\documentclass[10pt,journal,final]{IEEEtran}
\usepackage{amsmath,amsfonts}
\usepackage{amsthm}
\usepackage{amssymb}
\usepackage{algorithmic}
\usepackage{algorithm}
\usepackage{array}
\usepackage{mathrsfs}
\usepackage[caption=false,font=footnotesize,labelfont=rm,textfont=rm]{subfig}
\usepackage{textcomp}
\usepackage{stfloats}
\usepackage{url}
\usepackage{mathtools}
\usepackage{verbatim}
\usepackage{graphicx}
\usepackage{cite}
\usepackage{autobreak}
\usepackage{nicematrix}
\usepackage{arydshln}
\usepackage{latexsym}
\usepackage{booktabs}
\usepackage{nomencl}
\makenomenclature
\usepackage{subeqnarray}
\usepackage{cases}
\allowdisplaybreaks[4]

\usepackage{etoolbox}
\renewcommand\nomgroup[1]{%
	\item[\bfseries
	\ifstrequal{#1}{A}{Abbreviations}{%
	\ifstrequal{#1}{B}{Indices and Sets}{%
	\ifstrequal{#1}{C}{Parameters}{%
    \ifstrequal{#1}{D}{Variables}{%
    \ifstrequal{#1}{E}{Functions}{}}}}}%
]}

\begin{document}

\title{Privacy-Preserving Peer-to-Peer Energy Trading via Hybrid Secure Computations}
\vspace{-1cm}
\author{Junhong~Liu,~\IEEEmembership{Student~Member,~IEEE},~Qinfei~Long,~\IEEEmembership{Student~Member,~IEEE},~Rong-Peng~Liu,~\IEEEmembership{Member,~IEEE},\\~Wenjie~Liu,~\IEEEmembership{Member,~IEEE},~Xin~Cui,~\IEEEmembership{Student~Member,~IEEE},~and~Yunhe~Hou,~\IEEEmembership{Senior~Member,~IEEE}
\thanks{This work was supported in part by the National Natural Science Foundation of China (NSFC) under Grant 52177118, in part by the HUST-State Grid Future of Grid Institute under Grant 521209200014, and in part by the Research Grants Council of Hong Kong under Grant GRF 17209419. \emph{(Corresponding author: Yunhe Hou.)}} 
\thanks{Junhong Liu, Qinfei Long, Xin Cui, and Yunhe Hou are with the Department of Electrical and Electronic Engineering, The University of Hong Kong, Hong Kong SAR, China (e-mail: {jhliu, qflong, cuixin, yhhou}@eee.hku.hk).}
\thanks{Rong-Peng Liu is with the Department of Electrical and Computer Engineering, McGill University, Montréal, QC H3A 0E9, Canada (e-mail: rpliu@eee.hku.hk).}
\thanks{Wenjie Liu is with the Department of Mechanical and Automation Engineering, The Chinese University of Hong Kong, Hong Kong SAR, China (e-mail: wenjieliu@cuhk.edu.hk).} \vspace{-1.5em}}

\markboth{}%
{Shell \MakeLowercase{\textit{et al.}}: A Sample Article Using IEEEtran.cls for IEEE Journals}
\maketitle	
\begin{abstract}
The massive integration of uncertain distributed renewable energy resources into power systems raises power imbalance concerns. Peer-to-peer (P2P) energy trading provides a promising way to balance the prosumers' volatile energy power generation and demands locally. Particularly, to protect the privacy of prosumers, distributed P2P energy trading is broadly advocated. However, severe privacy leakage issues can emerge in the realistic fully distributed P2P energy trading paradigm. Meanwhile, in this paradigm, two-party and multi-party computations coexist, challenging the naive privacy-preserving techniques. To tackle privacy leakage issues arising from the fully distributed P2P energy trading, this paper proposes a privacy-preserving approach via hybrid secure computations. A secure multi-party computation mechanism consisting of offline and online phases is developed to ensure the security of shared data by leveraging the tailored secret sharing method. In addition, the Paillier encryption method based on the Chinese Remainder Theorem is proposed for both the secure two-party computation and the offline phase of the multi-party computation. The random encryption coefficient is designed to enhance the security of the two-party computation and simultaneously guarantee the convergence of the distributed optimization. The feasible range for the encryption coefficient is derived with a strict mathematical proof. Numerical simulations demonstrate the exactness, effectiveness, and scalability of the proposed privacy-preserving approach.
\end{abstract}


\vspace{-1.5em}

\mbox{}
\nomenclature[A]{\(Dec/Enc\)}{Decryption/Encryption}
\nomenclature[A]{\(Cmp\)}{Computation}
\nomenclature[A]{\(Cond.\)}{Condition}
\nomenclature[A]{\(No.\)}{Number}
\nomenclature[A]{\(PK/SK\)}{Public/Private key}
\nomenclature[A]{\(mod\)}{Modulo operation}

\nomenclature[B]{\( i,j,t\)}{Index of nodes/lines/agents}
\nomenclature[B]{\( k\)}{Index of iterations}
\nomenclature[B]{\({\cal H}_i\)}{Set of parent nodes for agent $i$}
\nomenclature[B]{\({\cal C}_i\)}{Set of child nodes for agent $i$}
\nomenclature[B]{\({\cal N}_p\)}{Set of total prosumers}
\nomenclature[B]{\({\cal S}_i\)}{Set of P2P trading partners for agent $i$}
\nomenclature[B]{\({\cal N}_b/{\cal N}_s\)}{Set of buyers/sellers}
\nomenclature[B]{\({\mathbb R}^D\)}{$D$-dimensional Euclidean space}
\nomenclature[B]{\({\mathbb Z}_n^*\)}{Set of the integers modulo $n$}

\nomenclature[C]{\( {\cal A}_i / {\cal B}_i\)}{Vector of global/local constraints of agent $i$}
\nomenclature[C]{\(\omega_b/\omega_s\)}{Price for buying/selling energy from/to the utility company}
\nomenclature[C]{\(\alpha_i,\beta_i\)}{Parameters of utility function of agent $i$}
\nomenclature[C]{\(\varepsilon_i\)}{Unit discomfort cost for load deviation of agent $i$}
\nomenclature[C]{\(p_{i,d}\)}{Preferred load demand of agent $i$}
\nomenclature[C]{\(p_i^{\min}/p_i^{\max}\)}{Lower/Upper bound of nodal active power injection of agent $i$}
\nomenclature[C]{\(q_i^{\min}/q_i^{\max}\)}{Lower/Upper bound of nodal reactive power injection of agent $i$}
\nomenclature[C]{\(v_i^{\min}/v_i^{\max}\)}{Lower/Upper bound of squared voltage magnitude of agent $i$}
\nomenclature[C]{\(P_i^{\min}/P_i^{\max}\)}{Lower/Upper bound of active power flow on the line connected by agent $i$ and its parent node}
\nomenclature[C]{\(Q_i^{\min}/Q_i^{\max}\)}{Lower/Upper bound of reactive power flow on the line connected by agent $i$ and its parent node}
\nomenclature[C]{\(\delta_i/\rho_i\)}{Smoothness/Convexity value of utility function of agent $i$}
\nomenclature[C]{\(R_i/X_i\)}{Resistance/Reactance of the line connected by agent $i$ and its parent node}
\nomenclature[C]{\(\xi_{i,a}/\xi_{i,a}^{\dagger}\)}{Varying/Initial step size of updating dual variables for global constraints of agent $i$}
\nomenclature[C]{\(\xi_{i,b}\)}{Varying step size of updating dual variables for local constraints of agent $i$}	
\nomenclature[C]{\(r_i\)}{Random encryption coefficient of agent $i$ in secure two-party computation}
\nomenclature[C]{\(\eta_i/\Breve{\eta}_i^{\dagger}\)}{Varying/Initial penalty parameter of agent $i$}
\nomenclature[C]{\(\mathcal{R}_i\)}{Secret random variable of agent $i$ in secure multi-party computation}
\nomenclature[C]{\(\Omega\)}{Sum of secret random variables for agent $i$ in secure multi-party computation}
\nomenclature[C]{\(\varphi_{i,m}\)}{Random variables of agent $i$ in secure multi-party computation}
\nomenclature[C]{\(\mathscr{G}_i(Z_i)/\mathcal{P}_i(Z_i)\)}{One/Sum of split shares of/for agent $i$ in secure multi-party computation}
\nomenclature[C]{\(\mu_i\)}{Varying step size of updating primal variables for agent $i$}
\nomenclature[C]{\(\tau\)}{Number of reserved decimal fraction digits in CRT-Paillier encryption}
\nomenclature[C]{\(d, d_1, d_2\)}{Plaintexts in CRT-Paillier encryption}
\nomenclature[C]{\(Z^*\)}{Upper bound of designed integer domain in CRT-Paillier encryption}
\nomenclature[C]{\(a_i/b_i\)}{Vector of constants in global/local constraints of agent $i$}

\nomenclature[D]{\( {\mathbf{\Phi}_i}\)}{Vector of primal variables of agent $i$}	
\nomenclature[D]{\(e_{i,j}\)}{Amount of traded energy between agents $i$ and $j$}
\nomenclature[D]{\(p_i/q_i\)}{Active/Reactive power injection of agent $i$}
\nomenclature[D]{\(v_i\)}{Squared voltage magnitude of agent $i$}
\nomenclature[D]{\(P_i/Q_i\)}{Active/Reactive power flow over the line connected by agent $i$ and its parent node}
\nomenclature[D]{\(\lambda_{i,\centerdot},\underline{\lambda_{i,\centerdot}},\overline{\lambda_{i,\centerdot}}\)}{Dual variables of agent $i$}
\nomenclature[D]{\(\boldsymbol{\lambda}_{i,a}\)}{Vector of dual variables for global constraints of agent $i$}
\nomenclature[D]{\(\boldsymbol{\lambda}_{i,b}\)}{Vector of dual variables for local constraints of agent $i$}
\nomenclature[D]{\(\gamma_p^k/\gamma_d^k\)}{Primal/Dual residual}
\nomenclature[E]{\(\nabla_\centerdot\)}{First-order partial differentiation with regard to the variable: $\centerdot$}
\nomenclature[E]{\(\mathcal{L}_i\)}{Augmented Lagrangian function of agent $i$}
\nomenclature[E]{\(\sigma_{max}(\centerdot)/\sigma_{min}(\centerdot)\)}{Maximum/Minimum singular value of the matrix: $\centerdot$}
\nomenclature[E]{\(\mathcal{G}_i\)}{Individual cost function of agent $i$}
\printnomenclature[2.2cm]

\section{Introduction}
\IEEEPARstart{D}{istributed} renewable energy resources, such as solar panels and wind generators, are extensively deployed in power systems to mitigate carbon emissions. However, the high penetration of uncertain renewable energy resources incurs potential troubles for the system operation, such as power imbalance\cite{shair2021power}. To alleviate these side effects, peer-to-peer (P2P) energy trading can be a promising mechanism that stimulates prosumers to actively trade their excessive renewable energy generation and thus helps balance their uncertain energy power locally\cite{morstyn2018using}. Particularly, the distributed P2P energy trading paradigm gains increasing popularity due to its ability to preserve  individual information for prosumers, such as private parameters of the utility function \cite{guo2021online,li2018distributed,zhong2020cooperative,li2022two}.

However, in reality, prosumers involved in the distributed P2P energy trading are required to share their partial data with multiple parties, i.e., neighbors, trading partners, and possibly a coordinator\cite{ullah2021peer,li2018distributed},  and this data-sharing mechanism raises privacy concerns. The shared data, especially the active power injection, contains individual sensitive energy usage profiles, which can be at risk of being directly leaked. As recently reported in \cite{rouf2012neighborhood}, variables and gradients exchanged among prosumers can be further exploited by malicious adversaries to infer private parameters of the utility function, which undermines the traditional distributed P2P energy trading. Specifically, extraneous adversaries can eavesdrop the insecure communication channels and intercept the shared data. Even agents in the market, i.e., neighbors, trading partners, and the coordinator, can infer the private information of prosumers by utilizing the shared data\cite{lu2018privacy}. As reported in \cite{zhao2014achieving}, adversaries can learn about residents’ behaviors through edge detection methods, which include the location and number of people inside a house. This poses serious security hazards for prosumers in the P2P energy market. By leveraging this private information, malicious adversaries can create unfair market competitions and even design tailored attacks to destroy network infrastructures \cite{shilov2021privacy}. Therefore, the sound distributed P2P energy trading paradigm should implement robust privacy-preserving techniques to protect sensitive shared data and further prevent any leakage or inference of private information. By doing so, all participants in the P2P energy trading system can engage in fair and secure trading without fear of their privacy being compromised.

In the recent literature, several privacy-preserving methods are proposed to enhance the privacy of the distributed optimization, namely differential privacy, homomorphic encryption, and secret sharing methods. The differential privacy method achieves privacy preservation of personal information by obfuscating shared data through inserting additional noises \cite{dwork2014algorithmic,smith2021realistic,dvorkin2023differentially}. Specifically, privacy issues of the distributed optimal power flow (OPF) are investigated in\cite{liu2018mitigating}, and shared data are obfuscated by adding uniform distributed stochastic noises to prevent potential inferences. Likewise, the Laplacian distributed perturbations are added to shared data in the distributed OPF problems\cite{fioretto2019differential}. Moreover, to ensure the personalized feasibility of randomized OPF solutions, an innovative privacy-preserving mechanism is proposed in \cite{dvorkin2020differential} by carefully calibrating noises through enforced chance constraints on grid limits. Nevertheless, the differential privacy method still has the limitation of the trade-off between the solution optimality and privacy preservation for individual agents.

Alternatively, the homomorphic encryption method can simultaneously guarantee the solution optimality and privacy preservation\cite{yi2014homomorphic}. This kind of method, such as the standard Paillier cryptosystem \cite{o2008paillier}, is investigated to preserve privacy in the distributed economic dispatch problems \cite{wu2021privacy,yan2021distributed}. Similarly, it is also explored to improve the security of federated learning to detect false data injection attacks in power systems\cite{li2022detection}. Nevertheless, privacy-preserving schemes based on the standard Paillier encryption method are computationally demanding, limiting their applications to real-time large-scale scenarios\cite{guo2021online}. Meanwhile, the homomorphic encryption-based privacy-preserving methods typically assume the existence of a trusted third party in charge of arithmetic operations on the encrypted data\cite{wu2021privacy,yan2021distributed,li2022detection}. The existence of a third party brings heavier communication burdens since each agent needs to iteratively interact with this additional party\cite{zhu2021distributed}. Moreover, the encryption methods with a trusted third party fail to protect participants' privacy for the two-party computation since one of the two parties can inversely infer the other’s information according to the computational result \cite{zhang2018admm}. 

Another privacy-preserving technique is the secret sharing method\cite{karnin1983secret}, which can also simultaneously ensure the solution optimality and privacy preservation. Specifically, a privacy-preserving distributed optimization scheme is proposed based on the traditional secret sharing method without sacrificing the solution optimality \cite{tian2021fully}. Similarly, an accurate distributed privacy-preserving EV charging protocol based on the traditional secret sharing method is developed in \cite{huo2022distributed}. However, the traditional secret sharing-based privacy-preserving methods require frequent communications and computations at each iteration, increasing burdens for the distributed optimization. Moreover, the secret sharing method only works for multi-party computation, i.e., the total number of involved parties should be larger than two. However, for fully distributed P2P energy trading, two-party and multi-party computations coexist, which challenges the existing privacy-preserving frameworks based on the secret sharing method.

To address privacy leakage issues stemming from the fully distributed P2P energy trading, we propose a privacy-preserving approach via hybrid secure computations. This approach takes advantage of the homomorphic encryption and secret sharing methods and bypasses the limitations of the differential privacy method (i.e., the trade-off between the solution optimality and privacy preservation). Specifically, the Paillier encryption method based on Chinese Remainder Theorem (CRT), i.e., CRT-Paillier encryption method, is proposed for both secure two-party and multi-party computations. The main contributions are summarized as follows:
\begin{itemize}
\item{We propose a novel privacy-preserving approach via hybrid secure computations for the fully distributed P2P energy trading problem. This approach does not rely on the trusted third party and can simultaneously guarantee the solution optimality and privacy preservation for individual agents.}
\item{We ensure the computational efficiency and privacy preservation for the two-party computation by employing the CRT-Paillier encryption method and designing additional random encryption coefficients. We also develop a secure multi-party computation mechanism consisting of offline and online phases to tackle privacy leakage concerns by leveraging the tailored secret sharing method. The security of shared data in the offline phase is guaranteed by the CRT-Paillier method, and the online phase can ensure the safety of shared data in a communicationally and computationally lightweight manner.}

\item{We further guarantee the linear convergence of the distributed optimization by generating random encryption coefficients (for the secure two-party computation) from the theoretically derived feasible range.}
\end{itemize}

The remainder of this paper is organized as follows: Section II introduces the general centralized P2P energy trading problem, its fully distributed reformulations, and the privacy leakage risks. Section III proposes a privacy-preserving approach via hybrid secure computations for the reformulated P2P energy trading problem. Section IV demonstrates the effectiveness of the proposed approach with case studies. Section V draws the conclusions.

\section{Formulations of the P2P Energy Trading Problem}
\subsection{Centralized P2P Energy Trading Problem}
This paper considers the real-time radial distribution-level P2P energy trading problem, which is formulated as:
\begin{subequations}
\begin{align}
\min\limits_{\mathbf{\Phi}} \quad & \sum\limits_{i \in {\cal N}_p} \{ {\alpha _i}\sum\limits_{j \in {\cal S}_i} {e_{i,j}^2} + {\beta _i}\sum\limits_{j \in {{\cal S}_i}} e_{i,j} + {\varepsilon _i} (p_i - p_{i,d})^2 \notag\\
&+ {\omega _b}[ -p_i + \sum\limits_{j \in {\cal S}_i} e_{i,j}]^+ - {\omega _s}[p_i - \sum\limits_{j \in {\cal S}_i} e_{i,j}]^+ \} \label{ob} \\
\mathbf{s.t.} \quad &p_i^{\min } \le {p_i} \le p_i^{\max}, i \in {\cal N}_p \label{box1}\\
&q_i^{\min } \le {q_i} \le q_i^{\max}, i \in {\cal N}_p\\
&v_i^{\min} \le {v_i} \le v_i^{\max},i \in {\cal N}_p\\
&P_i^{\min} \le {P_i} \le P_i^{\max}, i \in {\cal N}_p\\
&Q_i^{\min} \le {Q_i} \le Q_i^{\max}, i \in {\cal N}_p\\
&e_{i,j} \le 0, e_{j,i} \ge 0, i \in {\cal N}_b, j \in {\cal S}_i\\
&e_{i,j} \ge 0, e_{j,i} \le 0, i \in {\cal N}_s, j \in {\cal S}_i \label{box2}\\
&e_{i,j} + e_{j,i} = 0 \label{p2p},i \in {\cal N}_p, j \in {\cal S}_i \\
&v_{t} - v_i - 2({R_i}{P_i} + {X_i}{Q_i}) = 0, i \in {\cal N}_p, t \in {\mathcal H}_i \label{gg1} \\
&\sum\limits_{j \in {\cal C}_i} P_j - P_i - p_i = 0,i \in {\cal N}_p \label{gg2}\\
&\sum\limits_{j \in {\cal C}_i} Q_j - Q_i - q_i = 0, i \in {\cal N}_p \label{gg3}\\
&p_i - \sum\limits_{j \in {\cal S}_i} e_{i,j} \le 0,  i \in {\cal N}_b \label{ban1}\\
&\sum\limits_{j \in {\cal S}_i} e_{i,j} - p_i \le 0, i \in {\cal N}_s, \label{ban2}
\end{align}
\end{subequations}
where $[\centerdot]^{+}$ is the non-negative operator, which is equivalent to $max\{\centerdot,0\}$. Each prosumer locates at each node (bus) in the power system, with the utility company located at the bus 0. ${\mathcal H}_i$ is the set of parent nodes for agent $i$, which normally contains one element except for the bus 0. The objective function \eqref{ob} is to minimize the total operation cost or, equivalently, maximize the social welfare, which contains four parts: i) the first two items in \eqref{ob} are the utility function of the traded energy (active power) through the P2P energy trading; ii) the third item in \eqref{ob} represents the discomfit cost of rescheduling the active power from the desired power injection $p _{i,d}$ to meet the constraints; iii) the fourth term models the cost of buying energy from the utility company; iv) the fifth term models the expense of selling amounts of energy to the utility company. Vector $\mathbf{\Phi}$ represents the collection of decision variables of all the agents. Specifically, for agent $i$, $\mathbf{\Phi}_i=\{{p _i}, {q _i}, {P _i}, {Q _i}, {v _i}, {e _{i,j}}\}$. Constraints \eqref{box1}-\eqref{box2} refer to the boundary constraints for the decision variables. \eqref{p2p} models the equality constraint of amounts of traded energy between agents $i$ and $j$, i.e., $e _{i,j}$ and $e _{j,i}$, where $j \in {\mathcal S}_i$. Constraints \eqref{gg1}-\eqref{gg3} represent the linearized Distflow (LinDistflow) model \cite{yeh2012adaptive}, which is reported a 1\% error compared to the Distflow model \cite{zhou2019hierarchical,ullah2021peer}. To take into account the impacts of power losses on the energy trading, an additional linear term based on the electrical distance can be incorporated into the objective function \eqref{ob} following the practice in \cite{paudel2020peerloss,ullah2021peer,liu2023online}. This practice will not alter the structure and properties of the problem, and the proposed privacy-preserving approach still applies to the problem considering power losses. Constraints \eqref{ban1} and \eqref{ban2} model the energy balance relations. Note that agent $i$ can only be a buyer or a seller at one-time slot, depending on its desired active power injection $p_{i,d}$. Specifically, if $p _{i,d} < 0$, agent $i$ is a buyer and belongs to set ${\mathcal N}_b$. Accordingly, for this case, the constraint \eqref{ban2} is inactive and vice versa. For the case $p _{i,d} = 0$, agent $i$ is no longer involved in the P2P energy trading, and its individual cost function is set to be a constant, i.e., 0.

\subsection{Fully Distributed P2P Energy Trading Problem}
The centralized P2P energy trading problem (1) can be equivalently decomposed into node-based sub-problems. For agent $i$, the decomposed sub-problem is denoted by:
\begin{subequations}
	\begin{align}
		\mathcal{P}_1: \quad \min\limits_{\mathbf{\Phi}_i} \quad &  \mathcal{G}_i(\mathbf{\Phi}_i)\\
		\mathbf{s.t.} \quad & \mathcal{A}_i \cdot \mathbf{\Phi}_i - a_i = 0 \label{l1} \\
		\quad & \mathcal{B}_i \cdot \mathbf{\Phi}_i -b_i \le 0, \label{l2}
	\end{align}
\end{subequations}
where $\mathbf{\Phi}_i \in \mathbb{R}^D$, $\mathcal{A}_i$, $\mathcal{B}_i\in \mathbb{R}^{C \times D}$ and $ a_i$, $b_i\in \mathbb{R}^C$. $D$ is the dimension of decision variables and $C$ is the number of corresponding constraints. $\mathcal{G}_i(\mathbf{\Phi}_i):\mathbb{R}^D \to \mathbb{R}$ is the individual cost function for agent $i$ decomposed from \eqref{ob}. The equality constraint \eqref{l1} refers to the original global constraints \eqref{p2p}-\eqref{gg3}, where agent $i$ needs to share their data with its neighbors and trading partners. The inequality constraints \eqref{l2} refer to the original local constraints \eqref{box1}-\eqref{box2}, and \eqref{ban1} or \eqref{ban2}, where only agent $i$'s data are involved.

{\bf Assumption 1:} Cost function $\mathcal{G}_i(\mathbf{\Phi}_i): \mathbb{R}^D \to \mathbb{R}$ is assumed to be  $\delta_i$-smooth and $\rho_i$-strongly-convex as follows:
\begin{subequations}
	\begin{align}
		||\nabla\mathcal{G}_i(\mathbf{\Phi}_i)-\nabla\mathcal{G}_i(\mathbf{\Phi}_i^*)|| &\le \delta_i||\mathbf{\Phi}_i-\mathbf{\Phi}_i^*|| \\
		(\mathbf{\Phi}_i-\mathbf{\Phi}_i^*)^T(\nabla\mathcal{G}_i(\mathbf{\Phi}_i)-\nabla\mathcal{G}_i(\mathbf{\Phi}_i^*)) &\ge \rho_i||\mathbf{\Phi}_i-\mathbf{\Phi}_i^*||^2. 
	\end{align}
\end{subequations}

{\bf Remark 1:} Since each agent $i$ can be either a seller or a buyer at one-time slot, either the forth or fifth term in \eqref{ob} for agent $i$ takes effects. For the case that agent $i$ is no longer involved in the P2P energy trading, its cost function is a constant. Overall, the cost function $\mathcal{G}_i(\mathbf{\Phi}_i)$ is smooth and convex, and this assumption can be satisfied.

The node-based optimization problem $\mathcal{P}_1$ can be further reformulated into an unconstrained augmented Lagrangian problem and solved iteratively in analytic forms by the primal-dual hybrid gradient (PDHG) method \cite{he2014convergence}. Specifically, the unconstrained augmented Lagrangian optimization problem for agent $i$, reformulated from $\mathcal{P}_1$, is denoted as $\mathcal{P}_2$:
\begin{subequations}
\begin{align}
\mathcal{P}_2:\quad  & \min\limits_{\mathbf{\Phi}_i}  \quad  \mathcal{L}_i \\
where \quad & \mathcal{L}_i= \mathcal{G}_i(\mathbf{\Phi}_i)  \notag\\
&+\boldsymbol{\lambda}_{i,a}\cdot(\mathcal{A}_i \cdot \mathbf{\Phi}_i -a_i)+\frac{\eta _i}{2}||\mathcal{A}_i \cdot \mathbf{\Phi}_i -a_i||^2\notag \\
&+\boldsymbol{\lambda}_{i,b}\cdot(\mathcal{B}_i \cdot \mathbf{\Phi}_i - b_i ), 
\end{align}
\end{subequations}
where $\boldsymbol{\lambda}_{i,a} \in \mathbb{R}^C$ and $\boldsymbol{\lambda}_{i,b} \in \mathbb{R}^C$, and $\eta _i$ is the scalar penalty parameter. The augmented Lagrangian function of the decomposed P2P energy trading problem, take buyer $i$ for instance, can thus be formulated as:
\begin{align}
    \mathcal{L}_i=&{\alpha _i}\sum\limits_{j \in {\cal S}_i} {e_{i,j}^2} + {\beta _i}\sum\limits_{j \in {{\cal S}_i}} e_{i,j} + {\varepsilon _i} (p_i - p_{i,d})^2 \notag\\
&+{\omega _b}[ -p_i + \sum\limits_{j \in {\cal S}_i} e_{i,j}]^+ \notag\\
&+\underline{\lambda_{i,p}}(p_i^{min}-p_i)+\overline{\lambda_{i,p}}(p_i-p_i^{max}) \notag\\
&+\underline{\lambda_{i,q}}(q_i^{min}-q_i)+\overline{\lambda_{i,q}}(q_i-q_i^{max}) \notag\\
&+\underline{\lambda_{i,v}}(v_i^{min}-v_i)+\overline{\lambda_{i,v}}(v_i-v_i^{max}) \notag\\
&+\underline{\lambda_{i,P}}(P_i^{min}-P_i)+\overline{\lambda_{i,P}}(P_i-P_i^{max}) \notag \\
&+\underline{\lambda_{i,Q}}(Q_i^{min}-Q_i)+\overline{\lambda_{i,Q}}(Q_i-Q_i^{max}) \notag \\
&+{\lambda_{i,e}}(e_{i,j})+{\lambda_{i,ep}}(p_i - \sum\limits_{j \in {\cal S}_i} e_{i,j}) \notag\\
&+{\lambda_{i,pp}}(\sum\limits_{j \in {\cal C}_i} P_j - P_i - p_i)+\frac{\eta_i}{2}(\sum\limits_{j \in {\cal C}_i} P_j - P_i - p_i)^2 \notag\\
&+{\lambda_{i,qq}}(\sum\limits_{j \in {\cal C}_i} Q_j - Q_i - q_i)+\frac{\eta_i}{2}(\sum\limits_{j \in {\cal C}_i} Q_j - Q_i - q_i)^2  \notag\\
&+{\lambda_{i,vv}}\{v_{t} - v_i - 2({R_i}{P_i} + {X_i}{Q_i})\} \notag\\
&+\frac{\eta_i}{2}\{v_{t} - v_i - 2({R_i}{P_i} + {X_i}{Q_i})\}^2 \notag\\
&+{\lambda_{i,ee}}(e_{i,j}+e_{j,i})+\frac{\eta_i}{2}(e_{i,j}+e_{j,i})^2. 

\end{align}
 We can further derive the first-order gradients of the augmented Lagrangian function for each agent with regard to the primal and dual variables. Following the PDHG method, the optimization problem $\mathcal{P}_2$ can be solved iteratively by:
\begin{subequations}
\begin{align}
\mathcal{P}_3: \quad &\mathbf{\Phi}_i^{k+1}=\mathbf{\Phi}_i^k-{\mu _i}{\nabla}_{\mathbf{\Phi}_i} {\mathcal L}_i(\mathbf{\Phi}_i^k,\boldsymbol{\lambda}_{i,a}^{k},\boldsymbol{\lambda} _{i,b}^{k}) \label{update1}\\
&\boldsymbol{\lambda}_{i,a}^{k+1}=\boldsymbol{\lambda} _{i,a}^{k}+{\xi _{i,a}}{\nabla}_{\boldsymbol{\lambda _{i,a}}} {\mathcal L}_i(\mathbf{\Phi}_i^{k+1},\boldsymbol{\lambda}_{i,a}^{k},\boldsymbol{\lambda} _{i,b}^{k}) \label{update2} \\
&\boldsymbol{\lambda} _{i,b}^{k+1}=\boldsymbol{\lambda} _{i,b}^{k}+{\xi _{i,b}}{\nabla}_{\boldsymbol{\lambda _{i,b}}} {\mathcal L}_i(\mathbf{\Phi}_i^{k+1},\boldsymbol{\lambda}_{i,a}^{k},\boldsymbol{\lambda} _{i,b}^{k}), \label{update3}
\end{align}
\end{subequations}
where ${\mu _i}$ is the step size for updating the primal variables. ${\xi _{i,a}}$ and ${\xi _{i,b}}$ are the step sizes for updating the dual variables. Vector $\boldsymbol{\lambda}_{i,a}=\{\lambda_{i,ee},\lambda_{i,vv}, \lambda_{i,pp},\lambda_{i,qq}\}$ and vector $\boldsymbol{\lambda}_{i,b}=\{\underline{\lambda_{i,p}},\overline{\lambda_{i,p}},\underline{\lambda_{i,q}},\overline{\lambda_{i,q}},\underline{\lambda_{i,v}},\overline{\lambda_{i,v}},\underline{\lambda_{i,P}},\overline{\lambda_{i,P}},\underline{\lambda_{i,Q}},\overline{\lambda_{i,Q}},\lambda_{i,e},\lambda_{i,ep}$\}. Note that the coordinator is not compulsory, enabling the P2P energy trading problem to be solved in a fully distributed manner. However, intermediate variables are inevitably to be shared among neighbors and potential trading partners, leading to potential privacy-leakage issues.

\subsection{Privacy Leakage of Distributed P2P Energy Trading}
In the P2P energy trading problem, the bus voltage magnitude, active/reactive power injections, and parameters of the utility function are defined as the private data for prosumers\cite{zhang2018enabling,dvorkin2020differential,mak2020privacy}. However, these data can be inferred by others in the distributed data-sharing scheme. For instance, according to the updating rules of $\mathcal {P}_3$ for buyer $i$, adversaries can establish the following equations:
\begin{subequations}
 \begin{align}
 	p_i^k =& \sum\limits_{j \in {\cal C}_i} P_j^k - P_i^k  \label{ref1} \\
 	q_i^k =& \sum\limits_{j \in {\cal C}_i} Q_j^k - Q_i^k   \label{ref2} \\
	e_{i,j}^{k+1} = & e_{i,j}^{k}-\mu_i\{2\alpha_i\cdot e_{i,j}^k+\beta_i+\omega _b+\lambda_{i,e}^k-\lambda_{i,ep}^k+\lambda_{i,ee}^k  \notag \\
		              &+\eta_i(e_{i,j}^k+e_{j,i}^k)\}, \label{ref3}      
\end{align}
\end{subequations}		    
where equations \eqref{ref1} and \eqref{ref2} are derived from constraints \eqref{gg2} and \eqref{gg3} at the $k$-th iteration. Equation \eqref{ref3} is derived from the updating rule of $e_{i,j}$ for prosumer $i$. Adversaries can collect the shared data, i.e., $P_i^k$, $Q_i^k$, $P_j^k$, and $Q_j^k$, to infer the active/reactive power injections of agent $i$, i.e., $p_i^k$ and $q_i^k$, at every iteration following equations \eqref{ref1} and \eqref{ref2}. The bus voltage magnitude of agent $i$, $v_i^k$, can be directly derived by adversaries since it is shared among the neighbors. Moreover, adversaries can collect the shared data through intercepting communication channels over multiple iterations to infer private parameters of the utility function. Under two mild assumptions, i.e., the worst scenario \cite{ryu2021privacy}, we show that private parameters of the utility function can be inferred by adversaries within only two iterations. Assume that all the variables, i.e., $e_{i,j}^0, e_{j,i}^0, \lambda_{i,e}^0, \lambda_{i,ep}^0, \lambda_{i,ee}^0 $, are initialized as zero, and thus adversaries can derive the following equations:
\begin{subequations}
\begin{align}
	  \beta_i=- \frac{e_{i,j}^1}{\mu_i}-\omega _b  \label{ref4}  
\end{align}
\begin{numcases}
      {\alpha_i=}     
      \frac{1}{2e_{i,j}^{1}}\{\frac{e_{i,j}^1-e_{i,j}^2}{\mu_i}-(\xi_{i,a}+\eta_i)\cdot (e_{i,j}^1+e_{j,i}^1) \notag\\
       \quad +\xi_{i,b}(p_i^{1}-e_{i,j}^{1}-\sum\limits_{l \in {\cal S}_i} e_{i,l}^{1})-\beta_i-\omega_b\}  {\color{blue}\label{ref5}} \\
       \notag \\
      \frac{1}{2e_{i,j}^{1}}\{\frac{e_{i,j}^1-e_{i,j}^2}{\mu_i}-\beta_i-\omega_b-\eta_i\cdot (e_{i,j}^1+e_{j,i}^1) \notag\\
       \quad +\xi_{i,b}(p_i^{1}-e_{i,j}^{1}-\sum\limits_{l \in {\cal S}_i} e_{i,l}^{1})\}, {\color{blue}\label{ref6}}
\end{numcases}
\end{subequations}
where equation \eqref{ref4} is derived from equation \eqref{ref3} at the 1-th iteration. Equations \eqref{ref5} and \eqref{ref6}, corresponding to problems $\mathcal P_3$ and $\mathcal P_4$, are both derived from equation \eqref{ref3} at the 1-th and 2-th iterations.  
Additionally, when the default system settings, i.e. $\omega_b, \mu_i, \xi_{i,a}, \xi_{i,b}$, are the same for all agents in the market, and known to potential adversaries, private parameters of utility function for agent $i$, i.e., $\alpha_i$ and $\beta_i$, can be uniquely inferred by adversaries. The case study in Section IV.C further demonstrates this. To address these issues and simultaneously ensure exact solutions to the fully distributed P2P energy trading problem, in the next section, we propose an efficient privacy-preserving approach via hybrid secure computations.

\section{Proposed Privacy-Preserving Approach}
Typically, there are two circumstances when solving the fully distributed P2P energy trading problem: two-party and multi-party computations. For the multi-party circumstance, it appears when one agent has more than one child agent (node) and each child agent needs to share its variables with the parent agent for computing gradients cooperatively. Specifically, for the computation related to the global constraints \eqref{gg2}-\eqref{gg3}, it is a multi-party cooperation. The two-party computation appears when each agent has exactly one child agent. Specifically, for the computation related to the global constraint \eqref{gg1} and reciprocal P2P energy constraint \eqref{p2p}, it is always a two-party cooperation. For these two circumstances, the difficulties of preserving the privacy of shared data are distinct. The fewer agents a computation involves, e.g., the addition or subtraction operations, the more difficult it becomes to preserve privacy. For instance, when only two agents exist, it is easy for either agent to infer the other's data since the shared data cannot be mixed up with others as in the multi-party computation. Based on these concerns, we propose a privacy-preserving approach via hybrid secure computations, where the trusted third party can be avoided. Its overall approach is shown in Fig. 1, which is based on the CRT-Paillier encryption method (for secure two-party and multi-party computations) and the tailored secret sharing method (for the secure multi-party computation). To guarantee the convergence, we further derive the feasible range for the random encryption coefficient required in the secure two-party computation. 

\begin{figure*}[!tbp]
	\centering
	\includegraphics[width=7.2in]{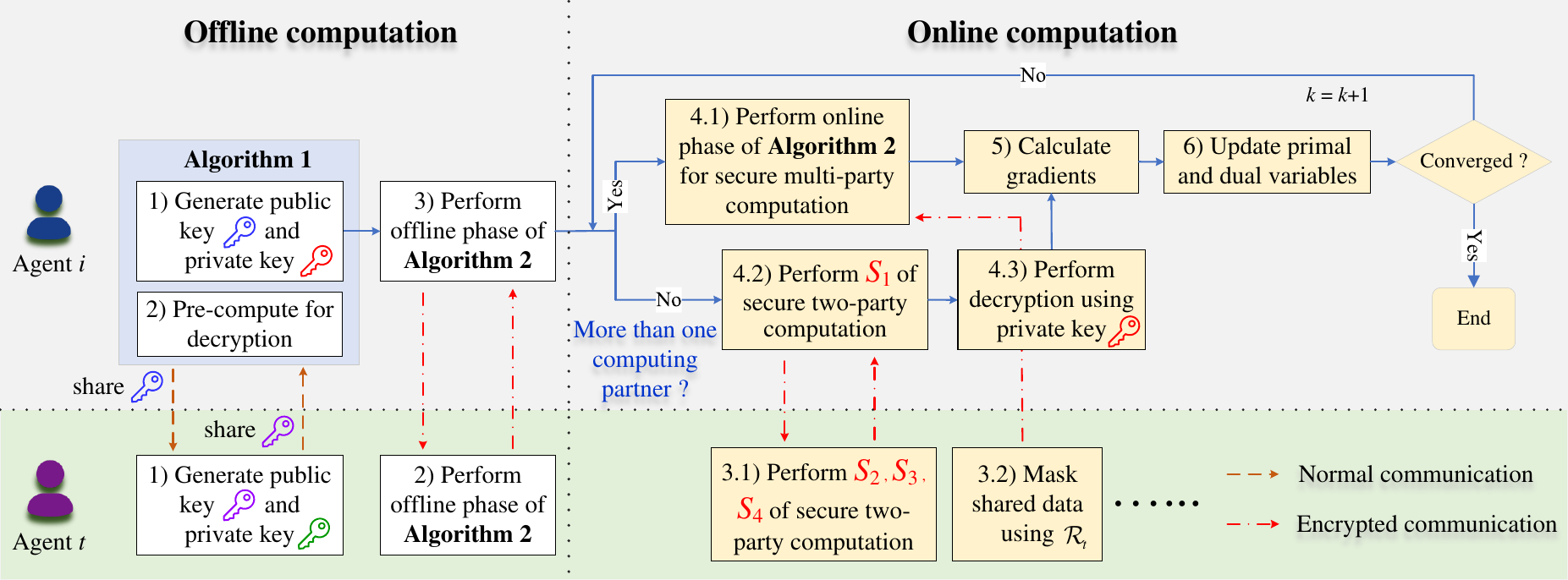}
	\caption{Proposed privacy-preserving approach via hybrid secure computations for the distributed P2P energy trading.}
    \label{fig1}
\end{figure*}
\vspace{-0.05cm}
\subsection{Secure Two-Party Computation}
For the secure two-party computation, only two agents are involved. For instance, if agent $i$ tends to update its voltage ${v_i}$, it needs to cooperate with its parent agent $t$, $t \in {\mathcal H}_i$, for computing the partial gradient as: 
\begin{align}
    {\nabla}_{v_i}=&-\lambda_{i,vv}^k-\eta_i(v_t^k-v_i^k-2(R_iP_i^k+X_iQ_i^k)) \notag\\
        &-\underline{\lambda_{i,v}^k}+\overline{\lambda_{i,v}^k}. \label{delta}
\end{align}

Equation \eqref{delta} is employed to update the primal variable $\mathbf{\Phi}_i^{k+1}$ in \eqref{update1} and the dual variable $\boldsymbol{\lambda} _{i,a}^{k+1}$ in \eqref{update2}. In order to compute the gradient ${\nabla}_{v_i}$ for agent $i$, agent $t$ needs to share (send) its own voltage data $v_t^k$ to agent $i$, which inevitably infringes on agent $t$'s privacy. To address this concern and simultaneously derive the correct gradients, we propose a secure two-party computation mechanism by leveraging the CRT-Paillier encryption system as in Algorithm 1. This algorithm mainly contains three steps, i.e., key generation, encryption, and decryption. Compared with the standard Paillier cryptosystem, the computational efficiency of the decryption step in Algorithm 1 is significantly improved. Specifically, for the decryption step $d \gets [\frac{((e)^{\pi} \text{ }mod\text{ } n^2)-1}{n}]\cdot \theta \text{ }mod\text{ } n $ in the standard one, the modular exponentiation operation $(e)^{\pi} \text{ }mod\text{ } n^2$  accounts for the major computational burden. To alleviate this, we thus propose the CRT-Paillier encryption cryptosystem, which divides the modular exponentiation operation into the sub-space of the modulus $n^2$, i.e., $ p^2 \text{ and } q^2$, and then combines the results as in $Dec(SK, e)$\cite{paillier1999public}. Besides, to support the encryption of floating point data, floating point numbers are mapped to integers by multiplying the coefficient $10^{\tau}$, where $\tau$ denotes the number of reserved decimal fraction digits. To support the encryption of negative data, the integer domain $(0, {Z^*}]$, where $ {Z^*} \ge {3\cdot max({\mathbb{Z}_{n}^*})}$ and set ${\mathbb{Z}_{n}^*}$ is defined by the length of keys, is divided into three intervals: $(0, \frac{{Z^*}}{3})$ for positive integers, $(\frac{2{Z^*}}{3}, {Z^*}]$ for negative integers, and $(\frac{{Z^*}}{3}, \frac{2{Z^*}}{3}]$ for detecting overflows arising from the addition and multiplication operations. When data ${D}$ is negative, it is mapped to a positive integer by ${D}^*={D}+{Z^*}$. We refer interested readers to \cite{paillier1999public,ogunseyi2020fast} for more details.  

\begin{algorithm}[!htbp]
\caption{CRT-Paillier Encryption Cryptosystem.}\label{alg1}
\begin{algorithmic}
\STATE 
\STATE {\textit{KeyGen}}
\STATE \hspace{0.5cm}$ \textbf{Generate} \text{ primes } p \text{ and } q \text{, where}$ gcd(pq,(p-1)(q-1))=1
\STATE \hspace{0.5cm}$ n \gets pq, \text{ and } \pi \gets (p-1)(q-1) $
\STATE \hspace{0.5cm}$ \theta \gets [(p-1)(q-1)]^{-1} \text{ }mod\text{ } n $
\STATE \hspace{0.5cm}$ PK \gets (n,n+1), SK \gets (\pi, \theta ) $
\STATE \hspace{0.5cm}\textbf{return}  $PK,  SK$
\STATE {\textit{Enc}}$(PK, d)$
\STATE \hspace{0.5cm}$ {D} \gets \lfloor10^\tau d\rfloor \in {\mathbb Z}_n^*$
\STATE \hspace{0.5cm}$ \textbf{Generate} \text{ random integer } r \in {\mathbb Z}_n^*$
\STATE \hspace{0.5cm}$ e \gets (1+n{D})\cdot r^n \text{ }mod\text{ } n^2 $
\STATE \hspace{0.5cm}\textbf{return}  $e$
\STATE {\textit{Dec}}$(SK, e)$
\STATE \hspace{0.5cm}$ {E}_p \gets \frac{(e^{p-1} \text{ }mod\text{ } p^2)-1}{p} \Gamma_p \text{ }mod\text{ } p$
\STATE \hspace{0.5cm}$ {E}_q \gets \frac{(e^{q-1} \text{ }mod\text{ } q^2)-1}{q} \Gamma_q \text{ }mod\text{ } q$
\STATE \hspace{0.5cm}$ {D} \gets {E}_q+{[(q^{-1}\text{ }mod\text{ } p})({E}_p-{E}_q){\text{ }mod\text{ }p}]\cdot q$
\STATE \hspace{0.5cm}$ d \gets {D}/10^\tau$
\STATE \hspace{0.5cm}$ \textbf{Pre-compute: } $
\STATE \hspace{1.5cm}$ \Gamma_p  \gets [\frac{((n+1)^{p-1} \text{ }mod\text{ } p^2)-1}{p}]^{-1} \text{ }mod\text{ } p$
\STATE \hspace{1.5cm}$ \Gamma_q  \gets [\frac{((n+1)^{q-1} \text{ }mod\text{ } q^2)-1}{q}]^{-1} \text{ }mod\text{ } q$
\STATE \hspace{0.5cm}\textbf{return} $d$
\end{algorithmic}
\end{algorithm}

\vspace{-0.2cm}
\begin{subequations}
	\begin{align}
		Dec(Enc(d_1)\cdot Enc(d_2) \text{ }mod\text{ } n^2)=&(d_1 + d_2)\text{ }mod\text{ }n \label{homo1}\\
		Dec(Enc(d_1)^{d_2}  \text{ }mod\text{ }n^2) =&(d_1 \cdot d_2)\text{ }mod \text{ }n. \label{homo2}
	\end{align}
\end{subequations}

\begin{figure}[!hbp]
	\centering
	\includegraphics[width=3.5in]{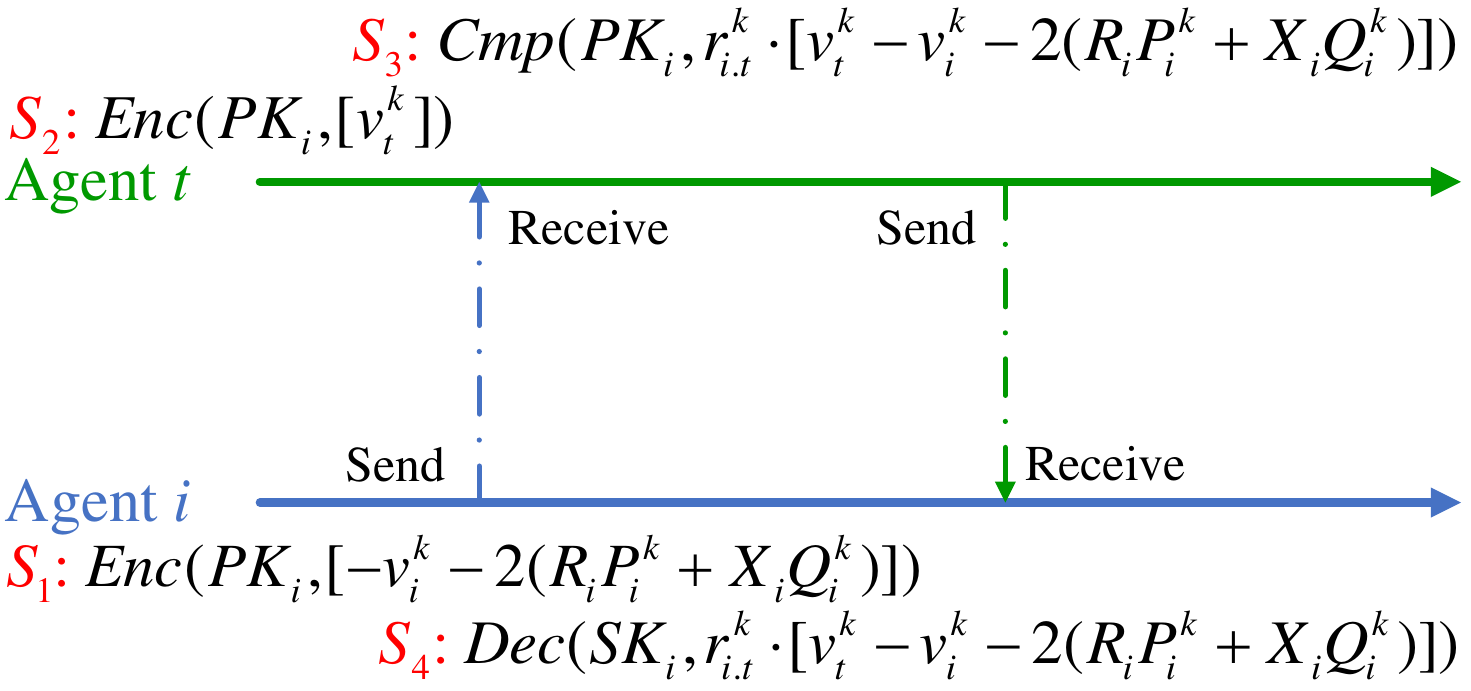}
	\caption{Secure two-party encrypted computation mechanism.}
	\label{fig_enc}
\end{figure}

The proposed CRT-Paillier encryption cryptosystem also preserves homomorphic properties as denoted by \eqref{homo1} and \eqref{homo2}. Based on these properties, we design the secure two-party computation mechanism as illustrated in Fig. \ref{fig_enc}. Agent $i$ first encrypts its own data as $S_1$ and then sends its public key $PK_i$ and the encrypted data to agent $t$. Agent $t$ encrypts its own data using agent $i$'s public key $PK_i$ as $S_2$ and then performs the addition operation directly on the encrypted ciphertext domain. To prohibit agent $i$ from inferring information by utilizing the shared data of agent $t$, a random encryption coefficient $r_{i,t}^k$ is generated by agent $t$ and multiplied by the sum of the data as $S_3$. When agent $i$ receives the encrypted data and decrypts it using its own secret key $SK_i$ as $S_4$, agent $i$ obtains the sum encrypted by the random encryption coefficient. Therefore, both agents have no idea of the other's precise data, but one agent obtains the encrypted sum of data from the two agents. The remaining task is to make sure that the random encryption coefficient has no impact on the optimal solutions and convergence of the distributed optimization. To this end, we derived its feasible range in Section III.C.

\subsection{Secure Multi-Party Computation}
The number of agents in the multi-party computation is larger than two. However, the proposed secure two-party computation mechanism cannot be extended to secure multi-party computation. Specifically, when agent $i$ has two child agents, i.e., agents 1 and 2, updating the reactive power for agent $i$ is a multi-party computation as:
\begin{align}
        {\nabla}_{q_i}=&-\lambda_{i,qq}^k-\eta_i((Q_{1}^k+Q_{2}^k)-Q_i^k-q_i^k)  \notag\\  \label{mu0}
        &-\underline{\lambda_{i,q}^k}+\overline{\lambda_{i,q}^k}. 
\end{align}
By applying the proposed secure two-party computation mechanism, we can obtain the following formula:
\begin{align}
        {\nabla}_{q_i}^{'}=&-\lambda_{i,qq}^k-\eta_i^{'}[r_{i,1}^k(Q_{1}^k+\frac{-Q_i^k-q_i^k}{2}) \notag \\
        &+r_{i,2}^k(Q_{2}^k+\frac{-Q_i^k-q_i^k}{2})]-\underline{\lambda_{i,q}^k}+\overline{\lambda_{i,q}^k}. 
\end{align}
The introduction of two random encryption coefficients $r_{i,1}^k$ and $r_{i,2}^k$ causes deviations from the original value, i.e., ${\nabla}_{q_i}^{'}-{\nabla}_{q_i} =(-\eta_i^{'}r_{i,1}^k+\eta_i)(Q_{1}^k+\frac{-Q_i^k-q_i^k}{2})+(-\eta_i^{'}r_{i,2}^k+\eta_i)(Q_{2}^k+\frac{-Q_i^k-q_i^k}{2})$. If agent $i$ has only the child agent 1, the deviation can be eliminated by designing $ \eta_i^{'}r_{i,1}^k=\eta_i$. However, when two child agents exist, the deviation can only be eliminated when $r_{i,1}^k=r_{i,2}^k=\frac{\eta_i}{\eta_i^{'}}$, which requires additional consensus protocols between agents 1 and 2, conflicting with the randomness of the encryption coefficients. Without such specific protocols, the deviation aggregates when the number of child agents increases, causing convergence problems for the distributed optimization. Alternatively, the multi-key encryption method can support the multi-party computation, nevertheless, demanding a trusted third party and suffering from the high computational complexity \cite{marcolla2022survey}. To address this concern, we propose an exact yet lightweight secure multi-party computation mechanism, as in Algorithm 2, by leveraging a tailored secret sharing method\cite{tjell2020private}. The proposed secure multi-party computation mechanism composes of the offline and online phases, which only brings trivial computational and communicational costs for the distributed optimization. Suppose agent $c$ is deriving its personal gradient values through the secure multi-party computation. Thus, total $m$, $m=|\mathcal{C}_c|+1$, agents are involved in this computation. Specifically, for the offline phase, each agent $i$, $i \in \mathcal{C}_c \cup {c}$, generates its own secret random variable $\mathcal{R}_i$ and splits the variable into $n$, $n \ge m$, shares, from $\mathscr{G}_i (Z_1)$ to $\mathscr{G}_i (Z_n)$ as:
\begin{subequations}
    \begin{align}
     \mathscr{G}_i(Z_1) = &\mathcal{R}_i + \varphi_{i,1} Z_1+...+\varphi_{i,m-1} Z_1^{m-1} \label{sp1}\\
     \mathscr{G}_i(Z_2) = &\mathcal{R}_i + \varphi_{i,1} Z_2+...+\varphi_{i,m-1} Z_2^{m-1} \\
        &  \vdots          \notag    \\
     \mathscr{G}_i(Z_n) = &\mathcal{R}_i + \varphi_{i,1} Z_n+...+\varphi_{i,m-1} Z_n^{m-1}, \label{sp2}
\end{align}
\end{subequations}
where $Z_1$, ..., $Z_{n}$ are positive integers, each corresponding to one agent. $\varphi_{i,1}$, ..., $\varphi_{i,m-1}$ are random variables generated by agent $i$. Each agent $i$ needs to send the split share $\mathscr{G}_i(Z_j)$ to agent $j$, $j \in \mathcal{C}_c \cup {c}$ and $j \ne i$, through the secure communication channel. To achieve secure data sharing, the CRT-Paillier cryptosystem is employed again. After receiving the split shares, each agent $j$ is able to compute the sum of split shares from all the $m$ agents by:
    \begin{align}
     &\mathcal{P}_j(Z_j)= \sum\limits_{i=1}^m(\mathscr{G}_i(Z_j)) \notag  \\ 
     &=\sum\limits_{i=1}^m(\mathcal{R}_i)+\sum\limits_{i=1}^m(\varphi_{i,1})Z_j+...+\sum\limits_{i=1}^m(\varphi_{i,m-1}) Z_j^{m-1}.   \label{sp3} 
\end{align}
Then each agent $j$, $j \in \mathcal{C}_c$, sends the sum of split shares, $\mathcal{P}_j(Z_j)$, through the secure communication channel to agent $c$. Note that agent $c$ is different from the trusted third party\cite{wu2021privacy,yan2021distributed,li2022detection}, since agent $c$ is the agent requiring the gradient values for updating individual primal and dual variables. Specifically, each agent encrypts the sum of split shares using the public key provided by agent $c$. When agent $c$ receives the encrypted sum, its true value can be retrieved by decryption. The sum of random variables from all agents can be reconstructed by agent $c$ using the Lagrangian interpolation method:
\begin{align}
     \Omega=&\sum\limits_{j=1}^m[\mathcal{P}_j(Z_j)\cdot\prod\limits_{h=1, h\neq j}^{m}\frac{Z_h}{Z_h-Z_j}]  \notag \\ \label{sp4}
     =&\sum\limits_{i=1}^m(\mathcal{R}_i).
\end{align}

 During the online phase, child agents of agent $c$, i.e., agent $j$ and $j \in \mathcal{C}_c$, mask their shared variables at iteration $k$ as:
\begin{align}
    \hat{\mathbf{\Phi}}_{j}^k=\mathbf{\Phi}_{j}^k+\mathcal{R}_j. \label{sp5}
\end{align}
Accordingly, agent $c$ can retrieve the true value of the sum of shared variables by subtracting $\Omega$:
\begin{align}
    \sum\limits_{j \in \mathcal{C}_c}({\mathbf{\Phi}}_{j}^k)=&\sum\limits_{j \in \mathcal{C}_c}({\hat{\mathbf{\Phi}}_{j}^k})+\mathcal{R}_c-\Omega. \label{sp6}
\end{align}
\vspace{-0.2cm}
\begin{algorithm}[!htbp]
\caption{Secure Multi-party Computation Mechanism.}\label{alg2}
\begin{algorithmic}
\STATE 
\STATE {\textsc{Offline Phase for agent} $i$, $i \in \mathcal{C}_c \cup {c}$}
\STATE \hspace{0.5cm}$ \textbf{Generate} \text{ secret random variable } \mathcal{R}_i$
\STATE \hspace{0.5cm}$ \textbf{Split } \mathcal{R}_i \text{ into $n$ shares following } \eqref{sp1} \text{ to } \eqref{sp2}$
\STATE \hspace{0.5cm}$ \textbf{Encrypt } \text{share } \mathscr{G}_i(Z_j) \text{ using agent } j \text{ 's public key } PK_j$
\STATE \hspace{0.5cm}$ \text{Send } Enc(PK_j,\mathscr{G}_i(Z_j)) \text{ to agent } j $
\STATE \hspace{0.5cm}$ \text{Receive all } Enc(PK_i,\mathscr{G}_j(Z_i)) \text{ from other agents} $
\STATE \hspace{0.5cm}$ \textbf{Decrypt } Enc(PK_i,\mathscr{G}_j(Z_i)) \text{ using private key } SK_i$
\STATE \hspace{0.5cm}$ \textbf{Compute } \text{the sum of split shares } \mathcal{P}_i(Z_i) \text{ following } \eqref{sp3}$
\STATE \hspace{0.5cm}$ \textbf{if} \text{ agent } i  \text{ is agent $c$:} $
\STATE \hspace{0.8cm}$ \text{Receive } Enc(PK_i,\mathcal{P}_j(Z_j)) \text{ from other agents} $
\STATE \hspace{0.8cm}$ \textbf{Decrypt } Enc(PK_i,\mathcal{P}_j(Z_j)) \text{ using private key } SK_i$
\STATE \hspace{0.8cm}$ \textbf{Reconstruct} \text{ } \Omega  \text{ following } \eqref{sp4}$
\STATE \hspace{0.5cm}$ \textbf{else} \text{:} $
\STATE \hspace{0.8cm}$ \textbf{Encrypt } \mathcal{P}_i(Z_i) \text{ using agent $c$} \text{'s public key}$
\STATE \hspace{0.8cm}$ \text{Send } Enc(PK_j,\mathcal{P}_i(Z_i)) \text{ to agent $c$} $
\STATE {\textsc{Online Phase for agent} $i$, $i \in \mathcal{C}_c \cup {c}$}
\STATE \hspace{0.5cm}$ \textbf{if} \text{ agent } i  \text{ is agent $c$:} $
\STATE \hspace{0.8cm}$ \text{Receive } \text{the masked values from other agents} $
\STATE \hspace{0.8cm}$ \textbf{Retrieve} \text{ the sum by subtracting } \Omega  \text{ following } \eqref{sp6}$
\STATE \hspace{0.5cm}$ \textbf{else} \text{:} $
\STATE \hspace{0.8cm}$ \textbf{Mask } \text{the shared variable with } \mathcal{R}_i\text{ following } \eqref{sp5}$
\STATE \hspace{0.8cm}$ \text{Send } \text{the masked value to agent $c$}$
\end{algorithmic}
\end{algorithm}
\vspace{-0.4cm}
\subsection{Convergence Analysis}
The secure multi-party computation is exact and does not impact the convergence of the distributed primal-dual hybrid gradient algorithm. However, the random encryption coefficient in the secure two-party computation does impact the convergence. We thus provide theoretical bounds for the random encryption coefficient $r_i$, $r_i >0$, for prosumer $i$ to ensure the convergence. Let $\Breve{\eta}_i^{\dagger}$ and $\xi_{i,a}^{\dagger}$ be initial values of the penalty parameter and step size for updating dual variables pertinent to the global constraints \eqref{l1}. The initialized values are constants for prosumer $i$ during the iterative updates.

{\bf Assumption 2:} The initial penalty parameter $\Breve{\eta}_i^{\dagger}$ is larger than the initial step size $\xi_{i,a}^{\dagger}$, which holds that $\Breve{\eta}_i^{\dagger}-\xi_{i,a}^{\dagger}>0$.

With {\bf Assumption 2}, the feasible range for the encryption coefficient $r_i$ can be derived as follows:

{\bf Theorem 1:} Let $0< k_{i,1}=\frac{1-\mu_i\delta_i}{\mu_i(\Breve{\eta}_i^{\dagger}-\xi_{i,a}^{\dagger})\cdot\sigma_{max}^2(\mathcal{A}_i)} $. To guarantee the linear convergence of the distributed optimization, the encryption coefficient $r_i$ and step size $\xi_{i,b} $ for prosumer $i$ should satisfy one of the following conditions:
\begin{subequations}
\begin{align}
 &{\bf Cond. 1:} \quad r_i \le min(k_{i,1},k_{i,2}), \quad \xi_{i,b}<\frac{\rho_i}{\sigma_{max}^2(\mathcal{B}_i)} \label{k0}\\
   &\quad \quad   k_{i,2}=\frac{\rho_i-\xi_{i,b}\cdot\sigma_{max}^2(\mathcal{B}_i)}{\xi_{i,a}^{\dagger}\cdot\sigma_{max}^2(\mathcal{A}_i)-(\Breve{\eta}_i^{\dagger}-\xi_{i,a}^{\dagger})\cdot\sigma_{min}^2(\mathcal{A}_i)} \label{k1}\\
 &when: \notag\\
 &\quad \quad  0< \xi_{i,a}^{\dagger}\cdot\sigma_{max}^2(\mathcal{A}_i)-(\Breve{\eta}_i^{\dagger}-\xi_{i,a}^{\dagger})\cdot\sigma_{min}^2(\mathcal{A}_i); \label{k1} \\
		&{\bf Cond. 2:} \quad k_{i,2} \le r_i \le k_{i,1}, \quad \xi_{i,b} > \frac{\rho_i}{\sigma_{max}^2(\mathcal{B}_i)} \label{k2}\\
		&\quad \quad   k_{i,2}=\frac{\xi_{i,b}\cdot\sigma_{max}^2(\mathcal{B}_i)-\rho_i}{(\Breve{\eta}_i^{\dagger}-\xi_{i,a}^{\dagger})\cdot\sigma_{min}^2(\mathcal{A}_i)-\xi_{i,a}^{\dagger}\cdot\sigma_{max}^2(\mathcal{A}_i)} \label{k1} \\
		&when: \notag\\
		&\quad \quad  0 > \xi_{i,a}^{\dagger}\cdot\sigma_{max}^2(\mathcal{A}_i)-(\Breve{\eta}_i^{\dagger}-\xi_{i,a}^{\dagger})\cdot\sigma_{min}^2(\mathcal{A}_i); \label{k3}\\
	&{\bf Cond. 3:} \quad r_i \le k_{i,1}, \quad \xi_{i,b} \le \frac{\rho_i}{\sigma_{max}^2(\mathcal{B}_i)} \label{k4}\\
	&when: \notag\\
	&\quad \quad  0 = \xi_{i,a}^{\dagger}\cdot\sigma_{max}^2(\mathcal{A}_i)-(\Breve{\eta}_i^{\dagger}-\xi_{i,a}^{\dagger})\cdot\sigma_{min}^2(\mathcal{A}_i). \label{k5}
\end{align}
\end{subequations}

Proof: We show the detailed proof in the Appendix.

For the secure two-party computation, agent $i$ derives its own feasible range for $r_i$ and shares this range with the other agent $j$ through the secure communication channel in advance. After receiving the feasible range, agent $j$ generates the random encryption coefficient $r_{i,j}$ within the feasible range, which is unknown to agent $i$ but can guarantee the convergence of the distributed optimization. Note that agent $i$ can share a random subset of its derived feasible range for $r_i$ to avoid potential privacy leakage from the feasible range.
\vspace{-0.1cm}
\subsection{Privacy Analysis}
Since the shared data is encrypted during the secure two-party and multi-party computations, extraneous eavesdroppers can only intercept encrypted ciphertexts of the shared data and cannot derive exact values of shared data without knowing the corresponding private key and the random variable $\mathcal{R}_i$. We thus focus on analyzing the privacy preservation of shared data against honest-but-curious agents under secure two-party and multi-party computations.

{\bf Theorem 2:} Suppose agents $i$ and $j$ are involved in the secure two-party computation. Agent $j$'s shared data $\boldsymbol{x}_j^k$ cannot be inferred by agent $i$ unless $\boldsymbol{x}_i^k=-\boldsymbol{x}_j^k$. 

Proof: We show the detailed proof in the Appendix.

{\bf Corollary 1:} For the secure multi-party computation, parent agent $i$ cannot derive exact values of the shared data from the child agent $j$, $ j \in \mathcal{C}_i$.

Proof: Following the reasoning of {\bf Theorem 2}, we show the detailed proof in the Appendix.
\section{Numerical Simulations}
\subsection{Simulation Setup}
The IEEE 15-bus distribution system \cite{ouali2020improved}, and the IEEE 34-bus, 69-bus, 94-bus, and 141-bus distribution systems \cite{zimmerman2010matpower,zhou2023fully} are employed to verify the optimality, efficiency, and scalability of the proposed approach. The code is implemented in GoLand 2022 on a computer with 6-core Intel (R) i5-10500 CPU@3.10GHz. The sizes of public and private keys for encryption methods are set to 128-bit to balance the security and computational efficiency following\cite{he2022privacy}. The real PV power generation and demand data are derived from \cite{lee2022grid,junhong2023}. Without losing generality\cite{ullah2021peer}, the following parameters are randomly generated for each prosumer: i) the parameters of the utility function for buyers are $\alpha _i$, $\alpha _i \in [0.01,0.1]$ ¢/kWh$^2$, and $\beta_i$, $\beta_i \in [1.0,3.0]$ ¢/kWh; ii) the corresponding parameters for sellers are $\alpha _i$, $\alpha _i \in [0.02,0.1]$ ¢/kWh$^2$, and $\beta_i$, $\beta_i \in [0.1,0.8]$ ¢/kWh; iii) the parameter for the discomfort cost function is $\epsilon _i$, $\epsilon _i \in [2.5,3.5]$ ¢/kWh$^2$. The penalty parameter $\Breve{\eta_i}^{\dagger}$ is set to $1.6$. The step size for updating primal variables, $\mu_i$, is set to $0.07$, and the step sizes for updating dual variables, i.e., $\xi_{i,a}^{\dagger}$ and $\xi_{i,b}$, are set to  $0.02$ and $0.015$, respectively. Accordingly, the feasible range for the random encryption coefficient can be obtained. Specifically, take buyers for instance, $\sigma_{min}(\mathcal{A}_i)=1.0$, $\sigma_{max}(\mathcal{A}_i)=1.4142$, and $\sigma_{max}(\mathcal{B}_i)=1.9021$. $\rho_i \ge min(2\alpha _i)=0.02$ and $\delta_i \le max(2\epsilon _i)=7.0$. Since $ \xi_{i,a}^{\dagger}\sigma_{max}^2(\mathcal{A}_i)-(\Breve{\eta}_i^{\dagger}-\xi_{i,a}^{\dagger})\sigma_{min}^2(\mathcal{A}_i)=-1.54 <0 $ and $ \xi_{i,b} > \frac{\rho_i}{\sigma_{max}^2(\mathcal{B}_i)}=0.00552$, ${\bf Cond.2}$ is satisfied. The feasible range is thus derived, i.e., $r_i \in [0.0223,2.3057]$. The subset of the feasible range for prosumer $i$, i.e., $[\underline{r_i},\overline{r_i}] \in [0.0223,2.3057]$, is sent to prosumer $j$, $j \in \mathcal{S}_i \cup \mathcal{A}_i \cup \mathcal{C}_i$.
\vspace{-0.5cm}
\subsection{Optimality and Convergence}
\begin{figure}[!tbp]
	\centering
	\includegraphics[width=1.0\linewidth]{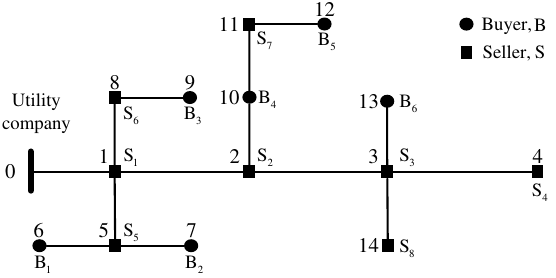}
	\caption{IEEE 15-bus distribution system with P2P energy trading \cite{lee2022grid,ullah2021peer}.}
	\label{fig3}
\end{figure}

\begin{figure}[!tbp]
\centering
\includegraphics[width=1.0\linewidth]{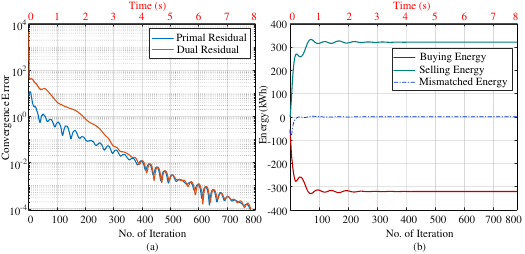}
\caption{ Performances of the convergence. (a) primal and dual residual errors. (b) buying and selling energy.}
\label{fig4}
\end{figure}

\begin{figure}[!tbp]
\centering
\includegraphics[width=1.0\linewidth]{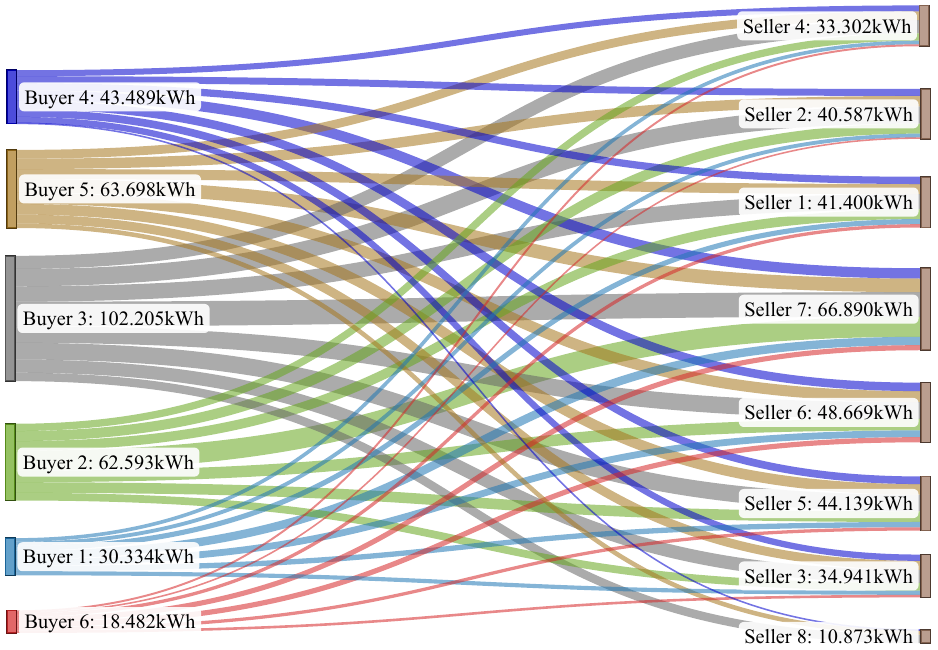}
\caption{Traded energy among different agents.}
\label{fig5}
\end{figure}

The optimality and convergence of the proposed hybrid privacy-preserving approach for the fully distributed P2P energy trading are verified on the 15-bus system, as shown in Fig. \ref{fig3}. The primal and dual residuals are denoted as $\gamma_p^k=||\sum\limits_{i \in {\cal N}_p}(\mathcal{A}_i \cdot \mathbf{\Phi}_i^{k} -a_i)||_2$ and $\gamma_d^k=||\sum\limits_{i \in {\cal N}_p}(\boldsymbol{\Phi}_i^{k+1}-\boldsymbol{\Phi}_i^{k})||_2$, respectively. We set $\gamma_p^k < 10^{-4}$ and  $\gamma_d^k < 10^{-4}$ as the stopping criteria. To demonstrate the optimality of the proposed distributed privacy-preserving approach (A3), we compare it with the centralized method (A1), distributed method without privacy preservation (A2), i.e., the PDHG method for directly solving $\mathcal{P}_3$. As in Table \ref{tab1}, the optimal amounts of total traded energy derived by A2 and A3 are equal, while A3 requires a slightly smaller number of iterations to converge than A2, which mainly arises from the effects of the random encryption coefficient in the secure two-party computation. As further illustrated in Fig. \ref{fig4}(a), the optimality gap of A3, i.e., $|(\mathcal{T}_{A3}-\mathcal{T}_{A1})/\mathcal{T}_{A1}|$, is less than $10^{-4}$ after 782 iterations. In addition, Fig. \ref{fig4}(b) shows that the mismatched energy derived from A3 tends to zero after 400 iterations or approximately 4 seconds. These testing results demonstrate the exactness, optimality, and convergence of the proposed hybrid privacy-preserving approach. Moreover, detailed amounts of traded energy among prosumers are illustrated in Fig. \ref{fig5}. Due to different parameter settings of utility functions, which indicate the prosumers' heterogeneous preferences, the amounts of traded energy among prosumers are diverse, e.g., buyer 1 exhibits a greater willingness to purchase energy from seller 7 in comparison to buyer 6. However, if the private parameters of buyer 1 are inferred by buyer 6 through intercepting shared data, buyer 6 can purposely take strategic measures, e.g., adjusting its own parameters to cause unfair market competition. As illustrated in Table \ref{tab2}, we show that the private data of buyer 1 can be easily inferred by other agents, including buyer 6, using equations \eqref{ref4} and \eqref{ref6}, and the inferred values of private parameters are equal to the true values. Therefore, to avoid such inferences and potential unfair manipulations, it is requisite to preserve the privacy of individual parameters against potential adversaries.
\begin{table}[htbp]
	\centering
	\caption{Optimality Comparisons}
	\begin{tabular}{cccc}
		\toprule  
		{\bf Methods}&{\bf Iterations}&{\bf Traded energy: $\mathcal T$(kWh)} \label{tab1}\\ 
		\cmidrule(r){2-4}
		A1 &-   &320.802 \\
    	A2 &784 &320.801 \\
	    A3 &782 &320.801 \\
		\bottomrule 
	\end{tabular}
\vspace{-0.5cm}
\end{table}

\begin{table}[!htbp]
	\centering
	\caption{Inference of Private Parameters of Utility Function}
	\begin{tabular}{cccccc}
		\toprule  
		&\bf True values &\bf Inferred values &\bf Relative error\label{tab2}\\ 
		\cmidrule(r){2-5}
		{$\alpha_i$}& 0.0297 &0.0297 &0.000\%\\
		{$\beta_i$}&1.0691 &1.0691 &0.000\%\\
		\bottomrule 
	\end{tabular}
\end{table}

\subsection{Performance of Privacy Preservation}
\begin{figure}[!tbp]
	\centering
	\includegraphics[width=1.0\linewidth]{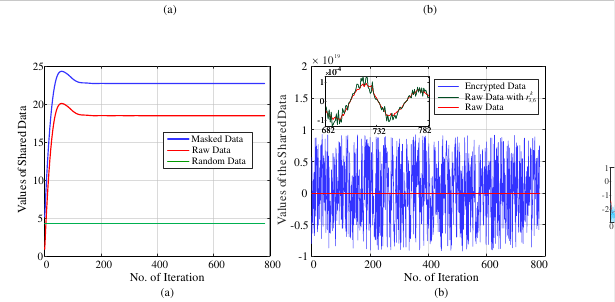}
	\caption{Performance of privacy preservation. (a) secure multi-party computation. (b) secure two-party computation.}
	\label{fig6}
\end{figure}

The privacy of data, including the decision variables and parameters of the utility function, is preserved by protecting the shared intermediate variables through hybrid secure two-party and multi-party computations. For illustrative purposes, the privacy-preserving performances for the shared data between seller 3 and buyer 6 are given. Seller 3 has one parent agent, i.e., seller 2, and three child agents, i.e., buyer 6, seller 4, and seller 8. Thus, the computation related to the global constraint \eqref{gg2} involves multiple parties. Fig. \ref{fig6}(a) demonstrates the evolution of masked shared data from buyer 6 to seller 3, i.e., $p_{6}^k+\mathcal{R}_6$ and $\mathcal{R}_6=4.2417$. $\mathcal{R}_6$ is the secret random variable for masking the shared data from buyer 6, i.e., $p_{6}^k$, which can preserve four decimal fraction digits. Since the secret random variable is only known to buyer 6, the shared data cannot be deciphered by seller 3 and other agents. Thus, the security of shared data is guaranteed. For the secure two-party computation related to the reciprocal energy constraint \eqref{p2p}, buyer 6 sends its personal encrypted value $Enc(PK_6,e_{6,3}^k)$ to seller 3. Seller 3 receives the value, encrypts its own value $e_{3,6}^k$ with buyer 6' public key, and then multiplies the encrypted ciphertext by a random encryption coefficient $r_{3,6}^k $, i.e., $Enc(PK_6, r_{3,6}^k(e_{6,3}^k+e_{3,6}^k))$. The coefficient $r_{3,6}^k$ is randomly generated within a shared subset of the feasible range from buyer 6, i.e., $r_{3,6}^k \in [0.5,1.5]$, and shared subsets from all the buyers are set to remain the same for illustrative purposes. As shown in Fig. \ref{fig6}(b), the encrypted ciphertext is at the order of $10^{19}$, while the raw data $(e_{6,3}^k+e_{3,6}^k)$ is at the order of $10^1$. The privacy of raw data is further preserved by the random encryption coefficient, i.e., $r_{3,6}^k$, as illustrated in the sub-figure of Fig. \ref{fig6}(b) for the last 100 iterations. The privacy-preserving performance of the random encryption coefficient can be further quantified by the multiplicative weight mechanism as in \cite{hardt2010multiplicative}, which is beyond the scope of this work. Therefore, the privacy of the shared data from seller 3 is protected from potential eavesdroppers and also buyer 6. This case fully demonstrates the effectiveness of the proposed hybrid approach for preserving the privacy of shared data in the two-party and multi-party computations.       
\vspace{-0.1cm}

\subsection{Computational Efficiency}
\begin{figure}[!tp]
	\centering
	\includegraphics[width=1.0\linewidth]{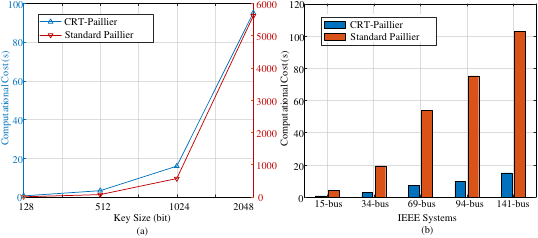}
	\caption{Average computational costs of two homomorphic encryption methods. (a) over different key sizes. (b) over different bus systems.}
	\label{fig7}
\end{figure}

This subsection further demonstrates the computational efficiency of the proposed privacy-preserving approach. The computational costs of the offline phase (of the proposed secure multi-party computation mechanism) for different systems are given in Table \ref{tab3}, and it indicates that the offline computational costs are trivial even for the 141-bus system. Moreover, computational and communicational comparisons between the multi-party computation protocols based on the traditional secret sharing method, denoted as M1, and the online phase (of the proposed secure multi-party computation mechanism), denoted as M2, are given in Table \ref{tab4}. It indicates that the online phase only brings lightweight communicational and computational costs for the distributed optimization. In fact, the computational cost of the proposed privacy-preserving approach via hybrid secure computations mainly comes from the online secure two-party computation. As shown in Table \ref{tab5}, the average computational cost of the CRT-Paillier method for each agent increases with the system size. For the 141-bus system, the computational cost of the CRT-Paillier method is $15.02s$, which meets the requirement of the real-time P2P energy trading with the time-scale  of 5 minutes or even less \cite{guo2021online}. In addition, we compare the proposed CRT-Paillier method with the standard Paillier method \cite{o2008paillier,wu2021privacy,yan2021distributed} over different key sizes and bus systems. As in Fig. \ref{fig7}(a), the computational costs of both methods increase exponentially with the key size. For the 15-bus system, the computational cost of the standard Paillier method is 9.26 times heavier than that of the CRT-Paillier method using the 128-bit key and 58.99 times heavier using the 2048-bit key. The standard Paillier method requires far more time than the proposed CRT-Paillier method for larger key sizes. The 128-bit key is thus more computationally efficient. Testing results in Fig. \ref{fig7}(b) indicate that the CRT-Paillier method also has far less computational burden than the standard Paillier method over different bus systems. For the standard Paillier method, the average computational cost for the 141-bus system is $102.82s$, which is 6.85 times higher than the CRT-Paillier method, confining its implementation in larger systems. By comparison, the proposed privacy-preserving approach enjoys good scalability.   
\begin{table}[!htbp]
	\centering
	\caption{Computational Costs of the Offline Phase}
	\begin{tabular}{cccccc}
		\toprule  
		 &15-bus&34-bus &69-bus&94-bus&141-bus \label{tab3}\\ 
		\cmidrule(r){2-6}
		{\bf Time (ms)}&9.99 &5.63&6.64&20.44&35.66 \\
		\bottomrule 
	\end{tabular}
 \vspace{-0.5cm}
\end{table}
\begin{table}[!htbp]
	\centering
	\caption{Complexity Comparisons of Multi-Party Computations}
	\begin{tabular}{cccc}
		\toprule  
		{\bf Methods}&{\bf Communication}&{\bf Computation} \label{tab4}\\ 
		\cmidrule(r){2-4}
		M1\cite{asmuth1983modular,tian2021fully,huo2022distributed} & $\mathcal{O}(m)$ & $\mathcal{O}(mlog^2m)$\\
		M2 & $\mathcal{O}(1)$ & $\mathcal{O}(1)$ \\
		\bottomrule 
	\end{tabular}
\end{table}
\begin{table}[!htp]
	\centering
	\caption{Average Computational Costs of CRT-Paillier Method}
	\begin{tabular}{cccccc}
		\toprule  
		 &15-bus&34-bus &69-bus&94-bus&141-bus \label{tab5}\\ 
		\cmidrule(r){2-6}
		{\bf Time (s)}&0.53 &2.91&7.19&9.63&15.02 \\
		\bottomrule 
	\end{tabular}
\end{table}
\section{Conclusions}
To address privacy leakage concerns in the fully distributed P2P energy trading problem caused by its data-sharing mechanism, we propose a privacy-preserving approach via hybrid secure computations. This approach takes advantage of the CRT-Paillier encryption and tailored secret sharing methods for secure two-party and multi-party computations, which can avoid the existence of a trusted third party. To enhance the security of shared data in the two-party computation, we design additional random encryption coefficients and provide a theoretically feasible range for these coefficients to guarantee convergence. Testing results demonstrate that the proposed hybrid privacy-preserving approach: i) ensures the optimal solutions and convergence of the problem; ii) preserves the privacy of shared data both in the two-party and multi-party computations; iii) has high computational efficiency and promising scalability.

{\appendix	
	\subsection{Proof of Theorem 1}
	We first derive the convergence conditions for the general case of problem $\mathcal{P}_3$ without considering hybrid encrypted computations, i.e., $r_i=1$. For this case, we extend the reasoning in \cite{alghunaim2020linear} to accommodate the vector of inequality constraints \eqref{l2} from problem $\mathcal{P}_1$. To accomplish this, three additional KKT conditions, \eqref{ap1}, \eqref{ap2}, and \eqref{ap3}, are adopted and relaxed for the derivations. Accordingly, more exact and complex convergence conditions are obtained for the general case of problem $\mathcal{P}_3$ compared to the existing literature\cite{alghunaim2020linear}. We further extend these obtained convergence conditions to derive the feasible range for the random encryption coefficient when hybrid secure computations are involved, i.e., $r_i$ is variable. Due to these differences from \cite{alghunaim2020linear}, we here provide detailed proofs for {\bf Theorem 1} specified in this paper. Suppose the pair of optimal solutions for agent $i$ is denoted as ($\mathbf{\Phi}_i^*, \boldsymbol{\lambda}_{i,a}^*,\boldsymbol{\lambda} _{i,b}^*$), and it should satisfy the following KKT conditions:
	\begin{subequations}
		\begin{align}
			&\nabla\mathcal{G}_\eta(\mathbf{\Phi}_i^*)+\mathcal{A}_i^T\cdot\boldsymbol{\lambda}_{i,a}^*+\mathcal{B}_i^T\cdot\boldsymbol{\lambda}_{i,b}^* =0 \\
			&\mathcal{A}_i\cdot\mathbf{\Phi}_i^*-a_i = 0 \\
			&\mathcal{B}_i\cdot\mathbf{\Phi}_i^*-b_i \le 0  \label{ap1} \\
			&\boldsymbol{\lambda}_{i,b}^* \ge 0 \label{ap2}\\
			&\boldsymbol{\lambda}_{i,b}^*\cdot (\mathcal{B}_i\mathbf{\Phi}_i^*-b_i) =0, \label{ap3}
		\end{align}
	\end{subequations}
	where $\mathcal{G}_\eta(\mathbf{\Phi}_i^*)=\mathcal{G}_i(\mathbf{\Phi}_i^*)+ \frac{\eta _i}{2}||\mathcal{A}_i \cdot \mathbf{\Phi}_i^* -a_i||^2$. Let $\mathbf{\tilde{\Phi}}_i=\mathbf{\Phi}_i-\mathbf{\Phi}_i^*$. $\boldsymbol{\tilde{\lambda}}_{i,b}=\boldsymbol{\lambda}_{i,b}-\boldsymbol{\lambda}_{i,b}^*$ and $\boldsymbol{\tilde{\lambda}}_{i,a}=\boldsymbol{\lambda}_{i,a}-\boldsymbol{\lambda}_{i,a}^*$ denote residuals for the primal and dual variables. Thus we can establish the following relations:
	\begin{subequations}
		\begin{align}
			\mathbf{\tilde{\Phi}}_i^{k+1}=&\mathbf{\tilde{\Phi}}_i^{k}-{\mu _i}[\nabla\mathcal{G}_\eta(\mathbf{\Phi}_i^{k})-\nabla\mathcal{G}_\eta(\mathbf{\Phi}_i^*)\notag\\
			&+\mathcal{A}_i^T\cdot\boldsymbol{\tilde{\lambda}}_{i,a}+\mathcal{B}_i^T\cdot\boldsymbol{\tilde{\lambda}}_{i,b}] \label{up1}\\
			\boldsymbol{\tilde{\lambda}}_{i,a}^{k+1}=&\boldsymbol{\tilde{\lambda}}_{i,a}^{k}+{\xi _{i,a}}(\mathcal{A}_i\mathbf{\tilde{\Phi}}_i^{k+1}) \label{up2}\\
			\boldsymbol{\tilde{\lambda}}_{i,b}^{k+1}\le &\boldsymbol{\tilde{\lambda}}_{i,b}^{k}+{\xi _{i,b}}(\mathcal{B}_i\mathbf{\tilde{\Phi}}_i^{k+1}). \label{up3}
		\end{align}
	\end{subequations}
	We can derive the following relations by taking the square of the above equations:
	\begin{subequations}
		\begin{align}
			||\mathbf{\tilde{\Phi}}_i^{k+1}||^2&=||\mathbf{\tilde{\Phi}}_i^{k}-{\mu _i}(\nabla\mathcal{G}_\eta(\mathbf{\Phi}_i^{k})-\nabla\mathcal{G}_\eta(\mathbf{\Phi}_i^*))||^2\notag\\
			&-2(\mathbf{\tilde{\Phi}}_i^{k}-{\mu_i}(\nabla\mathcal{G}_\eta(\mathbf{\Phi}_i^{k})-\nabla\mathcal{G}_\eta(\mathbf{\Phi}_i^*))) \notag\\
			& \cdot {\mu_i}(\mathcal{A}_i^T\cdot\boldsymbol{\tilde{\lambda}}_{i,a}+\mathcal{B}_i^T\cdot\boldsymbol{\tilde{\lambda}}_{i,b}) \notag\\
			&+{\mu_i^2}||\mathcal{A}_i^T\cdot\boldsymbol{\tilde{\lambda}}_{i,a}+\mathcal{B}_i^T\cdot\boldsymbol{\tilde{\lambda}}_{i,b}||^2 \\
			
			||\boldsymbol{\tilde{\lambda}}_{i,a}^{k+1}||^2&=||\boldsymbol{\tilde{\lambda}}_{i,a}^{k}||^2+2(\boldsymbol{\tilde{\lambda}}_{i,a}^{k})\cdot[{\xi _{i,a}}(\mathcal{A}_i\mathbf{\tilde{\Phi}}_i^{k+1})]\notag\\
			&+||{\xi _{i,a}}(\mathcal{A}_i\mathbf{\tilde{\Phi}}_i^{k+1})||^2 \label{ca2}\\
			||\boldsymbol{\tilde{\lambda}}_{i,b}^{k+1}||^2& \le ||\boldsymbol{\tilde{\lambda}}_{i,b}^{k}||^2+2(\boldsymbol{\tilde{\lambda}}_{i,b}^{k})\cdot[{\xi _{i,b}}(\mathcal{B}_i\mathbf{\tilde{\Phi}}_i^{k+1})]\notag\\
			&+||{\xi _{i,b}}(\mathcal{B}_i\mathbf{\tilde{\Phi}}_i^{k+1})||^2. \label{ca3}
		\end{align}
	\end{subequations}
	Furthermore, the following bounds can be derived:
	\begin{subequations}
		\begin{align}
			||{\xi _{i,a}}(\mathcal{A}_i\mathbf{\tilde{\Phi}}_i^{k})||^2 &\le {\xi _{i,a}^2}\cdot\sigma_{max}^2(\mathcal{A}_i)||\mathbf{\tilde{\Phi}}_i^{k}||^2\label{ca4} \\
			||{\xi _{i,b}}(\mathcal{B}_i\mathbf{\tilde{\Phi}}_i^{k})||^2 &\le {\xi _{i,b}^2}\cdot\sigma_{max}^2(\mathcal{B}_i)||\mathbf{\tilde{\Phi}}_i^{k}||^2.\label{ca5}
		\end{align}
	\end{subequations}
	Let $c_{\Phi}=1-\mu_i\xi_{i,a}\sigma_{max}^2(\mathcal{A}_i)-\mu_i\xi_{i,b}\sigma_{max}^2(\mathcal{B}_i)$. Replacing $\mathbf{\tilde{\Phi}}_i^{k+1}$ in \eqref{ca2} and \eqref{ca3} by its definition in \eqref{up1} and then multiplying $c_a={\mu_i}({\xi _{i,a}})^{-1}$ and $c_b={\mu_i}({\xi _{i,b}})^{-1}$, we can derive the following relations:
	\begin{align}
		&c_{\Phi}||\mathbf{\tilde{\Phi}}_i^{k+1}||^2+c_a||\boldsymbol{\tilde{\lambda}}_{i,a}^{k+1}||^2+c_b||\boldsymbol{\tilde{\lambda}}_{i,b}^{k+1}||^2 \le \notag\\
		&||\mathbf{\tilde{\Phi}}_i^{k}-{\mu _i}(\nabla\mathcal{G}_\eta(\mathbf{\Phi}_i^{k})-\nabla\mathcal{G}_\eta(\mathbf{\Phi}_i^*))||^2+c_a||\boldsymbol{\tilde{\lambda} }_{i,a}^{k}||^2+c_b||\boldsymbol{\tilde{\lambda}}_{i,b}^{k}||^2 \notag\\
		&-{\mu_i^2}||\mathcal{A}_i^T\cdot\boldsymbol{\tilde{\lambda}}_{i,a}^{k}+\mathcal{B}_i^T\cdot\boldsymbol{\tilde{\lambda}}_{i,b}^{k}||^2. \label{ca6}
	\end{align}
	Let $\mathcal{M}_i=[\mathcal{A}_i^T,\mathcal{B}_i^T]$ and $\boldsymbol{\lambda}_i^k=[\boldsymbol{\tilde{\lambda}}_{i,a}^{k};\boldsymbol{\tilde{\lambda} }_{i,b}^{k}]$, and the following relations can be obtained:
	\begin{align}
		||\mathcal{A}_i^T\cdot\boldsymbol{\tilde{\lambda}}_{i,a}^{k}+\mathcal{B}_i^T\cdot\boldsymbol{\tilde{\lambda}}_{i,b}^{k}||^2&=||\mathcal{M}_i \cdot \boldsymbol{\lambda}_i^k||^2 \notag\\ \label{ineq}
		&\ge {\sigma_{min}}^2(\mathcal{M}_i)||\boldsymbol{\lambda}_i^k||^2.
	\end{align}
	The inequality \eqref{ineq} can be obtained when $\boldsymbol{\lambda}_i^k$ is in the range space of $\mathcal{M}_i$, as proved by Lemma 1 and Lemma 2 in \cite{alghunaim2020linear}. Let $c_{\lambda}=[c_a, c_b]$, and the following relations can be derived:
	\begin{subequations}
		\begin{align}
			&c_{\Phi}||\mathbf{\tilde{\Phi}}_i^{k+1}||^2+c_{\lambda}||\boldsymbol{\lambda}_i^{k+1}||^2 \le \\ &||\mathbf{\tilde{\Phi}}_i^{k}-{\mu _i}(\nabla\mathcal{G}_\eta(\mathbf{\Phi}_i^{k})-\nabla\mathcal{G}_\eta(\mathbf{\Phi}_i^*))||^2 
			+ c_{\lambda}||\boldsymbol{\lambda}_i^{k}||^2 \notag\\
			& - c_{\lambda} \cdot \left [\begin{array}{cc}
				\mu_i\xi_{i,a}{\sigma_{min}}^2(\mathcal{M}_i) & 0 \notag \\
				0   & \mu_i\xi_{i,b}{\sigma_{min}}^2(\mathcal{M}_i)
			\end{array} \right ]||\boldsymbol{\lambda}_i^k||^2 \notag\\
			&=||\mathbf{\tilde{\Phi}}_i^{k}-{\mu _i}(\nabla\mathcal{G}_\eta(\mathbf{\Phi}_i^{k})-\nabla\mathcal{G}_\eta(\mathbf{\Phi}_i^*))||^2+ \\
			&c_{\lambda} \cdot \left [\begin{array}{cc}
				1-\mu_i\xi_{i,a}{\sigma_{min}}^2(\mathcal{M}_i) & 0  \notag\\
				0   & 1-\mu_i\xi_{i,b}{\sigma_{min}}^2(\mathcal{M}_i)
			\end{array} \right ]||\boldsymbol{\lambda}_i^k||^2.
		\end{align}
	\end{subequations}
	Function $\mathcal{G}_\eta(\mathbf{\Phi}_i)=\mathcal{G}_i(\mathbf{\Phi}_i)+ \frac{\eta _i}{2}||\mathcal{A}_i \cdot \mathbf{\Phi}_i -a_i||^2$ is $\delta_{\eta}$-smooth with $\delta_{\eta}=\delta_i+\eta_i\sigma_{max}^2(\mathcal{A}_i)$ and $\rho_{\eta}$-strongly-convex with $0 < \rho_{\eta} \le \delta_{\eta}$. Function $\mathcal{G}_\eta(\mathbf{\Phi}_i)$ holds the properties \cite{nesterov2003introductory}, i.e., $||\nabla\mathcal{G}_\eta(\mathbf{\Phi}_i)-\nabla\mathcal{G}_\eta(\mathbf{\Phi}_i^*)||^2 \le \delta_{\eta}\tilde{\mathbf{\Phi}}_i^T(\nabla\mathcal{G}_\eta(\mathbf{\Phi}_i)-\nabla\mathcal{G}_\eta(\mathbf{\Phi}_i^*)) $ and $(\mathbf{\Phi}_i-\mathbf{\Phi}_i^*)^T(\nabla\mathcal{G}_\eta(\mathbf{\Phi}_i)-\nabla\mathcal{G}_\eta(\mathbf{\Phi}_i^*)) \ge \rho_{\eta}||\mathbf{\Phi}_i-\mathbf{\Phi}_i^*||^2.$
	We thus can further derive the following relations:
	\begin{subequations}
		\begin{align}
			&||\mathbf{\tilde{\Phi}}_i^{k}-{\mu _i}(\nabla\mathcal{G}_\eta(\mathbf{\Phi}_i^{k})-\nabla\mathcal{G}_\eta(\mathbf{\Phi}_i^*))||^2 \le
			(\mathbf{\tilde{\Phi}}_i^{k})^2 \label{cc3}\\
			&+(\mu_i^2\delta_{\eta}-2\mu_i)(\mathbf{\tilde{\Phi}}_i^{k})^T(\nabla\mathcal{G}_\eta(\mathbf{\Phi}_i)-\nabla\mathcal{G}_\eta(\mathbf{\Phi}_i^*)) \notag\\
			& \le (\mathbf{\tilde{\Phi}}_i^{k})^2+(\mu_i^2\delta_{\eta}-2\mu_i)\rho_{\eta}||\mathbf{\tilde{\Phi}}_i^{k}||^2 \label{cc4}\\
			&=[1+(\mu_i^2\delta_{\eta}-2\mu_i)\rho_{\eta}] ||\mathbf{\tilde{\Phi}}_i^{k}||^2 \notag\\
			&=[\theta_ic_{\Phi}-(\rho_{\eta}-\theta_i(\xi_{i,a}\sigma_{max}^2(\mathcal{A}_i)+\xi_{i,b}\sigma_{max}^2(\mathcal{B}_i)))\mu_i]||\mathbf{\tilde{\Phi}}_i^{k}||^2 \notag\\
			&\le \theta_i c_\Phi||\mathbf{\tilde{\Phi}}_i^{k}||^2, \label{cc5}
		\end{align}
	\end{subequations}
	where $\theta_i=1+(\mu_i^2\delta_{\eta}-\mu_i)\rho_{\eta}$. The second inequality \eqref{cc4} is derived under the condition that $\mu_i\delta_{\eta}-2<0$. The third inequality \eqref{cc5} holds under the conditions that $\mu_i\delta_{\eta}-1<0 $ and $ \rho_{\eta}-\theta_i(\xi_{i,a}\sigma_{max}^2(\mathcal{A}_i)+\xi_{i,b}\sigma_{max}^2(\mathcal{B}_i)) \ge 0$. Since the condition, $\mu_i\delta_{\eta}-1<0$, is much stricter than the one, i.e., $\mu_i\delta_{\eta}-2<0$, the second inequality \eqref{cc4} can be thus derived and meanwhile it holds that $\theta_i<1$ when $\mu_i\delta_{\eta}-1<0$. We can conclude it as follows: 
	\begin{align}
		&c_{\Phi}||\mathbf{\tilde{\Phi}}_i^{k+1}||^2+c_{\lambda}||\boldsymbol{\lambda}_i^{k+1}||^2 \le \theta_i c_\Phi||\mathbf{\tilde{\Phi}}_i^{k}||^2+  \label{conver}\\
		&c_{\lambda} \cdot \left [\begin{array}{cc}
			1-\mu_i\xi_{i,a}{\sigma_{min}}^2(\mathcal{M}_i) & 0  \notag\\
			0   & 1-\mu_i\xi_{i,b}{\sigma_{min}}^2(\mathcal{M}_i)
		\end{array} \right ]||\boldsymbol{\lambda}_i^k||^2.
	\end{align}
	If $\mu_i\delta_{\eta}-1<0 $ and $ \rho_{\eta}-\xi_{i,a}\sigma_{max}^2(\mathcal{A}_i)-\xi_{i,b}\sigma_{max}^2(\mathcal{B}_i)\ge0$ are satisfied, the distributed optimization can achieve convergence with a linear rate of $l_i=max(\theta_i,1-\mu_i\xi_{i,a}{\sigma_{min}}^2(\mathcal{M}_i),1-\mu_i\xi_{i,b}{\sigma_{min}}^2(\mathcal{M}_i)) <1$ for each prosumer $i$.
	
	However, as in \eqref{update2} and \eqref{up2}, the newest primal variable $\mathbf{{\Phi}}_i^{k+1}$ is utilized to update the dual variable $\boldsymbol{{\lambda}}_{i,a}^{k+1}$, which indicates the encryption, encrypted computation, and decryption processes will be implemented again. To avoid such a huge computational burden, we further revise the updating process in \eqref{update2} to be non-incremental by using the old primal variable $\mathbf{{\Phi}}_i^{k}$. Note that the updating formula \eqref{update3} still keeps the same by adopting the newest primal variable $\mathbf{{\Phi}}_i^{k+1}$ since it represents the local constraints \eqref{l2} and does not need the data from other agents. The non-incremental forms are as follows:
	\begin{subequations}
		\begin{align}
			\mathcal{P}_4:  \mathbf{\Phi}_i^{k+1}=&\mathbf{\Phi}_i^k-{\mu _i}{\nabla}_{\mathbf{\Phi}_i} \Breve{\mathcal L}_i(\mathbf{\Phi}_i^k,\boldsymbol{\lambda}_{i,a}^{k},\boldsymbol{\lambda} _{i,b}^{k}) \label{up4}\\
			\boldsymbol{\lambda}_{i,a}^{k+1}=&\boldsymbol{\lambda} _{i,a}^{k}+\xi_{i,a}^{\dagger}\cdot r_i\cdot\textbf{}{\nabla}_{\boldsymbol{\lambda _{i,a}}} \Breve{\mathcal L}_i(\mathbf{\Phi}_i^{k},\boldsymbol{\lambda}_{i,a}^{k},\boldsymbol{\lambda} _{i,b}^{k}) \label{up5} \\
			\boldsymbol{\lambda} _{i,b}^{k+1}=&\boldsymbol{\lambda} _{i,b}^{k}+{\xi _{i,b}}{\nabla}_{\boldsymbol{\lambda _{i,b}}} \Breve{\mathcal L}_i(\mathbf{\Phi}_i^{k+1},\boldsymbol{\lambda}_{i,a}^{k},\boldsymbol{\lambda} _{i,b}^{k}). \label{up6}
		\end{align}
	\end{subequations}
	$\mathcal{P}_4$ can be reformulated into $\mathcal{P}_3$ under the condition that $\Breve{\mathcal L}_i(\mathbf{\Phi}_i^k,\boldsymbol{\lambda}_{i,a}^{k},\boldsymbol{\lambda}_{i,b}^{k})={\mathcal L}_i(\mathbf{\Phi}_i^k,\boldsymbol{\lambda}_{i,a}^{k},\boldsymbol{\lambda}_{i,b}^{k})+\frac{\xi_{i,a}^{\dagger}\cdot r_i}{2}||\mathcal{A}_i \cdot\mathbf{\Phi}_i^k- a_i||^2$. By letting $\boldsymbol{\Breve{\lambda}}_{i,a}^k=\boldsymbol{\lambda}_{i,a}^k+\xi_{i,a}^{\dagger}\cdot r_i\cdot(\mathcal{A}_i\cdot\mathbf{\Phi}_i^k-a_i)$ and adding $\xi_{i,a}^{\dagger}\cdot r_i\cdot(\mathcal{A}_i\cdot\mathbf{\Phi}_i^{k+1}-a_i)$ at both sides of \eqref{up5}. The derived convergence criteria in \eqref{conver} for $\mathcal{P}_3$ can be applied to $\mathcal{P}_4$ by defining  $\mathcal{G}_{\eta}(\mathbf{\Phi}_i)=\mathcal{G}_{\Breve{\eta}}(\mathbf{\Phi}_i)-\frac{\xi_{i,a}^{\dagger}\cdot r_i}{2}||\mathcal{A}_i \cdot \mathbf{\Phi}_i- a_i||^2=\mathcal{G}_i(\mathbf{\Phi}_i)+ \frac{\Breve{\eta}_i^{\dagger}\cdot r_i}{2}||\mathcal{A}_i \cdot \mathbf{\Phi}_i -a_i||^2-\frac{\xi_{i,a}^{\dagger}\cdot r_i}{2}||\mathcal{A}_i \cdot \mathbf{\Phi}_i- a_i||^2$. Under {\bf Assumption 2}, we can derive that $\eta_i=\Breve{\eta}_i^{\dagger}\cdot r_i-\xi_{i,a}^{\dagger}\cdot r_i>0$. Therefore, function $\mathcal{G}_{\eta}(\mathbf{\Phi}_i)$ is $\delta_{\eta}$-smooth with $\delta_{\eta}=\delta_i+(\Breve{\eta}_i^{\dagger}\cdot r_i-\xi_{i,a}^{\dagger}\cdot r_i)\cdot\sigma_{max}^2(\mathcal{A}_i)$ and $\rho_{\eta}$-strongly-convex with $\rho_{\eta}=\rho_i+(\Breve{\eta}_i^{\dagger}\cdot r_i-\xi_{i,a}^{\dagger}\cdot r_i)\cdot\sigma_{min}^2(\mathcal{A}_i)$. To guarantee the linear convergence of the distributed optimization, the encryption coefficient $r_i$ and step size $\xi_{i,b} $ should satisfy the conditions: $\mu_i[\delta_i+(\Breve{\eta}_i^{\dagger}\cdot r_i-\xi_{i,a}^{\dagger}\cdot r_i)\cdot\sigma_{max}^2(\mathcal{A}_i)]-1<0 \label{t1} $ and $ \rho_i+(\Breve{\eta}_i^{\dagger}\cdot r_i-\xi_{i,a}^{\dagger}\cdot r_i)\cdot\sigma_{min}^2(\mathcal{A}_i)-\xi_{i,a}^{\dagger}\cdot r_i\cdot\sigma_{max}^2(\mathcal{A}_i)-\xi_{i,b}\cdot\sigma_{max}^2(\mathcal{B}_i) \ge 0 \label{t2}$.
	The conditions for the encryption coefficient $r_i$ and stepsize $\xi_{i,b}$ can be further derived. When $\mu_i\delta_{\eta}-1<0 $ and $\delta_{i}<\delta_{\eta}$, it automatically holds that $1-\mu_i\delta_{i}>0$ and $k_{i,1}=\frac{1-\mu_i\delta_i}{\mu_i(\Breve{\eta}_i^{\dagger}-\xi_{i,a}^{\dagger})\cdot\sigma_{max}^2(\mathcal{A}_i)}>0$. There are thus three circumstances, i.e., $(\Breve{\eta}_i^{\dagger}\cdot r_i-\xi_{i,a}^{\dagger}\cdot r_i)\cdot\sigma_{min}^2(\mathcal{A}_i)-\xi_{i,a}^{\dagger}\cdot r_i\cdot\sigma_{max}^2(\mathcal{A}_i) < 0$, $(\Breve{\eta}_i^{\dagger}\cdot r_i-\xi_{i,a}^{\dagger}\cdot r_i)\cdot\sigma_{min}^2(\mathcal{A}_i)-\xi_{i,a}^{\dagger}\cdot r_i\cdot\sigma_{max}^2(\mathcal{A}_i) > 0$, and $(\Breve{\eta}_i^{\dagger}\cdot r_i-\xi_{i,a}^{\dagger}\cdot r_i)\cdot\sigma_{min}^2(\mathcal{A}_i)-\xi_{i,a}^{\dagger}\cdot r_i\cdot\sigma_{max}^2(\mathcal{A}_i) = 0 $, and they are concluded as ${\bf Cond.1}$, ${\bf Cond.2}$, and ${\bf Cond.3}$, respectively. This completes the proof of {\bf Theorem 1}. $\hfill\blacksquare$
	\subsection{Proof of Theorem 2}
	Suppose agent $i$ strictly follows the proposed secure two-party computation mechanism and collects data $\boldsymbol{x}_j \in \mathbb{R}^D$ from agent $j$ together with its own variable $\boldsymbol{x}_i\in \mathbb{R}^D $ for the updating process over $K$ iterations:
	\begin{subequations}
	\begin{align}
		\boldsymbol{y}_i^0=&r_{i,j}^0(\boldsymbol{x}_i^0+\boldsymbol{x}_j^0)  \\
		\boldsymbol{y}_i^k=&r_{i,j}^k(\boldsymbol{x}_i^k+\boldsymbol{x}_j^k)  \\
		&\vdots                  \notag \\
		\boldsymbol{y}_i^K=&r_{i,j}^K(\boldsymbol{x}_i^K+\boldsymbol{x}_j^K).  
	\end{align}
	\end{subequations}
	For agent $i$, $r_{i,j}^k$ and $\boldsymbol{x}_j^k (k=0,1,...,K)$ are unknown variables, while $\boldsymbol{y}_i^k$ and $\boldsymbol{x}_i^k (k=0,1,...,K)$ are known variables. Over $K$ iterations, agent $i$ can establish $D(K+1)$ equations with $(D+1)(K+1)$ unknown variables. Since the unknown variable outnumbers the established equation, agent $i$ cannot derive the exact values of shared data $\boldsymbol{x}_j \in \mathbb{R}^D$ by solving the equations. While the distributed optimization converges,  $\boldsymbol{x}_i^k+\boldsymbol{x}_j^k=0$. It holds that $\boldsymbol{y}_i^k=0$ for any encryption coefficient $r_{i,j}^k$. Agent $i$ can thus inversely obtain converged values of $\boldsymbol{x}_j \in \mathbb{R}^D$, but cannot infer intermediate values. This completes the proof of {\bf Theorem 2}.$\hfill\blacksquare$
    \subsection{Proof of Corollary 1}
	The following equations can be established for agent $i$ involved in the computation when collecting $\boldsymbol{\Phi}_j \in \mathbb{R}^D$ from agent $j$, $j \in \mathcal{C}_i$, over $K$ iterations:
	\begin{subequations}
	\begin{align}
		\boldsymbol{y}_i^0=&\sum\limits_{j \in \mathcal{C}_i}(\boldsymbol{\Phi}_{i,j}^0+\mathcal{R}_j)-\Omega  \\
		\boldsymbol{y}_i^1=&\sum\limits_{j \in \mathcal{C}_i}(\boldsymbol{\Phi}_{i,j}^1+\mathcal{R}_j)-\Omega  \\
		&\vdots                  \notag \\
		\boldsymbol{y}_i^K=&\sum\limits_{j \in \mathcal{C}_i}(\boldsymbol{\Phi}_{i,j}^K+\mathcal{R}_j)-\Omega.
	\end{align}
	\end{subequations}
	For agent $i$, $\mathcal{R}_j$ and $\boldsymbol{\Phi}_{i,j}^k (k=0,1,...,K)$ are unknown, but $\Omega$, $\boldsymbol{\Phi}_{i,j}^k+\mathcal{R}_j (k=0,1,...,K)$, and $\boldsymbol{y}_i^k (k=0,1,...,K)$ are known. Agent $i$ can establish $D(K+1)$ equations with at least $D(K+1)+1 $ unknown variables. Similarly, agent $i$ cannot infer exact values of the unknown variables from agent $j$, $j \in \mathcal{C}_i$, by solving the established equations. This completes the proof of {\bf Corollary 1}. $\hfill\blacksquare$
}

\ifCLASSOPTIONcaptionsoff
\newpage
\fi

\bibliographystyle{IEEEtran}
\bibliography{index}

\end{document}